\documentclass[pdflatex,sn-basic]{sn-jnl} 

\jyear{2021}%

\theoremstyle{thmstyleone}

\theoremstyle{thmstyletwo}

\theoremstyle{thmstylethree}

\usepackage{amssymb}
\usepackage{pifont}

\usepackage{amsmath}
\usepackage{lineno}
\usepackage{accsupp}
\usepackage{fancyhdr}
\usepackage{lastpage}
\usepackage{adjustbox}
\usepackage{tabularx}
\usepackage{makecell}
\usepackage{multirow}

\begin{document}

\title[Article Title]{Geostatistics in the presence of multivariate complexities: comparison of multi-Gaussian transforms}

\author*[1,2]{\fnm{Sultan} \sur{Abulkhair}}\email{sultan.abulkhair@adelaide.edu.au}

\author[1,2]{\fnm{Peter A.} \sur{Dowd}}\nomail

\author[1]{\fnm{Chaoshui} \sur{Xu}}\nomail

\affil[1]{\orgdiv{ARC Training Centre for Integrated Operations for Complex Resources}, \orgname{University of Adelaide}, \orgaddress{\city{Adelaide}, \country{Australia}}}

\affil[2]{\orgdiv{School of Civil, Environmental \& Mining Engineering}, \orgname{University of Adelaide}, \orgaddress{\city{Adelaide}, \country{Australia}}}

\abstract{One of the most challenging aspects of multivariate geostatistics is dealing with complex relationships between variables. Geostatistical co-simulation and spatial decorrelation methods, commonly used for modelling multiple variables, are ineffective in the presence of multivariate complexities. On the other hand, multi-Gaussian transforms are designed to deal with complex multivariate relationships, such as non-linearity, heteroscedasticity and geological constraints. These methods transform the variables into independent multi-Gaussian factors that can be individually simulated. This study compares the performance of the following multi-Gaussian transforms: rotation based iterative Gaussianisation, projection pursuit multivariate transform and flow transformation. Case studies with bivariate complexities are used to evaluate and compare the realisations of the transformed values. For this purpose, commonly used geostatistical validation metrics are applied, including multivariate normality tests, reproduction of bivariate relationships, and histogram and variogram validation. Based on most of the metrics, all three methods produced results of similar quality. The most obvious difference is the execution speed for forward and back transformation, for which flow transformation is much slower.}


\maketitle

\section{Introduction}\label{sec1}

Geostatistical conditional simulation often requires the modelling of multiple cross-correlated variables. For example, multivariate geostatistics is commonly applied to model ore grades and other mineral deposit variables \citep{bib1,bib2}. Among the Gaussian methods, there are three ways to perform this task \citep{bib3}. The first is to use conditional co-simulation algorithms such as sequential Gaussian co-simulation \citep{bib4} and turning bands co-simulation \citep{bib5}. The second approach is to use hierarchical co-simulation, which enables full, collocated and multi-collocated co-kriging to be applied \citep{bib6}. Finally, the third way is to transform variables into uncorrelated factors and simulate them individually.

Direct and hierarchical co-simulations use a linear model of co-regionalisation (LMC) \citep{bib1} or a Markov model \citep{bib7} to account for cross-correlations between variables. However, inference of cross-variograms can be challenging, whereas transformation to independent factors significantly simplifies multivariate modelling. For example, principal component analysis (PCA) and independent component analysis (ICA) have been applied in geostatistics to transform cross-correlated variables into independent orthogonal factors \citep{bib8,bib9}. However, the validity of extending PCA and ICA decorrelation to non-zero lag distances is an assumption. Alternatively, minimum/maximum autocorrelation factors (MAF) performs a double spectral decomposition at lag zero and a single non-zero lag to achieve spatial decorrelation \citep{bib10}. Other spatial decorrelation methods applied in geostatistics are uniformly weighted exhaustive diagonalisation with Gauss iterations and rotational joint diagonalisation \citep{bib11,bib12}.

Geostatistical co-simulation is also limited by the multi-Gaussianity assumption, which is often unrealistic as there may be complex multivariate relationships \citep{bib13}. LMC cannot include these complexities in a covariance matrix, and linear transformations (e.g., PCA, ICA and MAF) are only applicable to linear relationships. This motivated \cite{bib14} to apply stepwise conditional transformation (SCT) in the geostatistical framework to deal with multivariate complexities, such as non-linearity, heteroscedasticity and inequality constraints. In geostatistics, SCT was the earliest application of a multi-Gaussian transform (MGT). The idea behind MGT approaches is to simply transform variables into standard multivariate Gaussian distributions with zero correlation. SCT removes the complexities by transforming the original variables into multi-Gaussian factors. However, SCT cannot handle high-dimensional datasets and it requires extensive data cleaning and the order of the variables to be predefined \citep{bib3}. Recently, \cite{bib15} proposed a direct multivariate simulation based on SCT and demonstrated its applicability to a six-dimensional dataset with non-linear relationships.

Two of the most popular MGT methods in geostatistics are the projection pursuit multivariate transform (PPMT) \citep{bib13,bib16} and flow transformation, also known as flow anamorphosis (FA) \citep{bib17}. PPMT searches for the direction that has the maximum projection index \citep{bib18} and applies a normal score transformation in that direction. It is an iterative algorithm that transforms the data along projections based on the departure from Gaussianity. PPMT can be applied to a higher number of variables than SCT, it can work with smaller datasets and does not need the order of variables to be predefined \citep{bib13}. On the other hand, FA continuously deforms the original distribution into multi-Gaussian space using Lagrangian mechanics \citep{bib17}. Furthermore, its affine equivariance makes FA suitable for compositional data analysis \citep{bib19}, in which it can be paired with various log-ratio transforms \citep{bib20}.

A case study of the Gol-e-Gohar iron deposit by \cite{bib21} suggests that the combination of log-ratio+FA+MAF produces fewer artefacts during back transformations and provides better reproduction of compositional constraints than the combination of PPMT+MAF. In the case study, the additive log-ratio transformation was chained only with FA because of its affine equivariance. However, \cite{bib22} successfully chained PPMT with an isometric log-ratio transformation \citep{bib23} to reproduce the sum constraint. Similarly, additive and fraction ratios (i.e., not logarithmic) have been used on PPMT factors to reproduce sum and fractional constraints, also known as inequality constraints \citep{bib24}. Nevertheless, it is also essential to chain these methods with MAF or any other spatial decorrelation to ensure that factors are independent at non-zero lags. For example, the combination of PPMT+MAF shows significantly better variogram reproduction compared to the transformation by PPMT or by MAF \citep{bib25}.

This study compared the performance of PPMT, FA and a relatively new method to geostatistics, rotation based iterative Gaussianisation (RBIG) \citep{bib26}. RBIG is similar to PPMT, but it rotates the data using either PCA or ICA and applies the normal score transformation after each rotation. The following sections provide more details on selected methods together with a comprehensive comparison based on different metrics. The comparison is based on three bivariate case studies from undisclosed mining deposits with strong multivariate complexities. The results were carefully assessed using different statistical and qualitative metrics.

\section{Materials and methods}\label{sec2}

\subsection{Case studies with multivariate complexities}\label{subsec2.1}

\begin{figure}[ht]%
\centering
\includegraphics[width=1\textwidth]{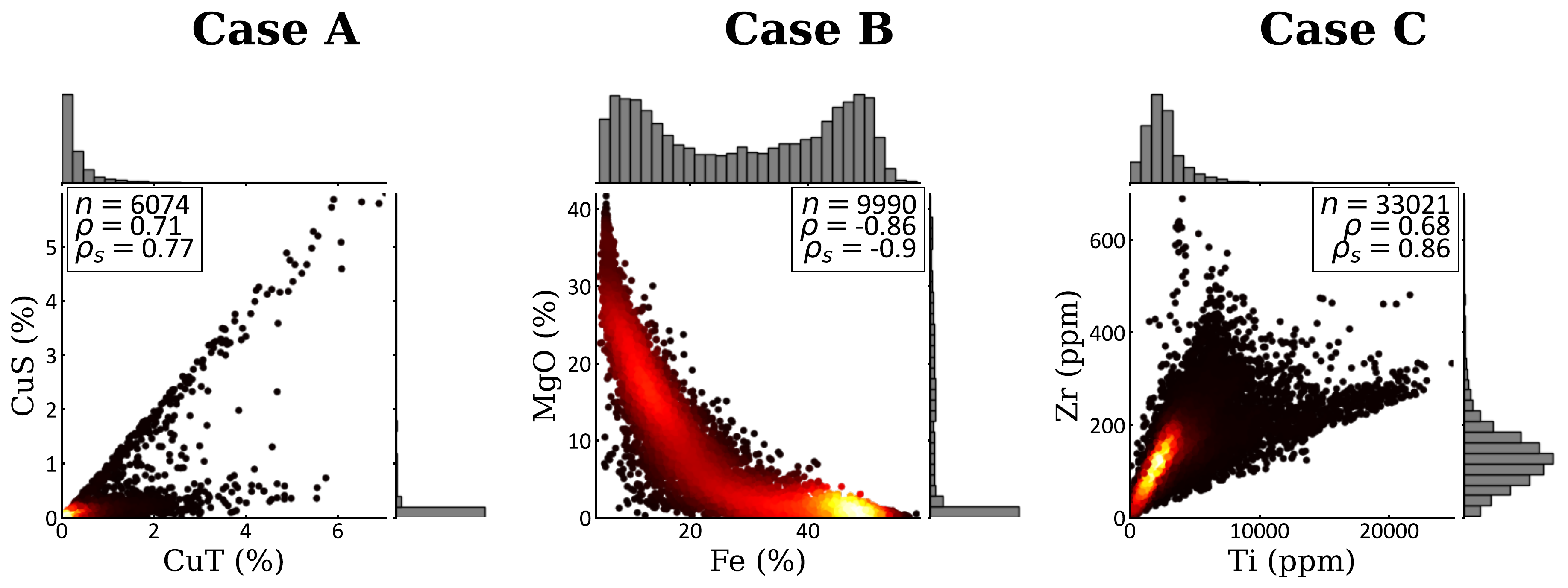}
\caption{Cross plots of three bivariate case studies with an inequality constraint (case A), non-linearity (case B) and heteroscedasticity (case C). Number of samples, Pearson $\rho$ and Spearman $\rho_S$ correlation coefficients are shown in text boxes.}\label{fig1}
\end{figure}

This paper applies selected methods to three bivariate case studies with complex relationships (Figure \ref{fig1}). These are confidential mining datasets and are, therefore, undisclosed. Case A consists of 6,074 drill hole samples for which there is an inequality constraint between total and soluble copper grades. The average horizontal spacing between samples is 123 m with a composite length of 2 m. This type of multivariate complexity occurs when one variable is a fraction of another variable, and there have been attempts to model this relationship \citep{bib28,bib24,bib27}. Inequality constraints can also occur in iron ore deposits between iron and elements such as silica and aluminium oxide when there is a linear inequality between variables \citep{bib29,bib30}.

For case B, there is a non-linearity between 9,990 iron and magnesium oxide samples. In this case, the sample points are irregularly spaced with an average of 4.2 m for horizontal and 2.4 m for vertical spacing. Non-linearities are not rare in drill hole datasets, and several geostatistical case studies have focussed on modelling such datasets \citep{bib14,bib13,bib16,bib15,bib25}. 

Finally, the case C data come from an underground mine with an average of 14.5 m spacing and a composite length of 2 m. It is a bivariate dataset with 33,021 samples of titanium and zirconium with a heteroscedastic relationship between them. Heteroscedasticity is similar to a linear relationship, expressed by a non-constant variance of one variable across the range of values of another. Heteroscedastic relationships has been reported in many multivariate geostatistical studies \citep{bib13,bib16,bib15,bib25}. However, case C demonstrates a more obvious heteroscedastic relationship with a moderately high correlation.

As all three cases are mining datasets prone to sampling irregularities, duplicates and outliers, they underwent careful data cleaning. In addition, cell declustering \citep{bib31} was applied for cases A and B. Cells of 85m × 85m × 85m were chosen for case A and 18m × 18m × 18m for case B after evaluating the effect of different cell sizes. As a result, mean and standard deviation values decreased after declustering for both cases. Cell declustering was not performed for case C because the sum of weights was not equal to the number of samples. Table \ref{tab1} shows the descriptive statistics for all three case studies. As multivariate complexities are involved, both Spearman and Pearson correlation coefficients are reported.

\begin{table}[ht]
\begin{center}
\caption{Descriptive statistics of three bivariate datasets}\label{tab1}%
\begin{adjustbox}{width=0.9\textwidth}
\begin{tabular}{@{}lcccccc@{}}
\toprule
\multirow{2}{*}{Parameter} & \multicolumn{2}{c}{Case A} & \multicolumn{2}{c}{Case B} & \multicolumn{2}{c}{Case B} \\
& CuT (\%) & CuS (\%) & Fe (\%) & MgO (\%) & Ti (ppm) & Zr (ppm) \\
\midrule
Mean & 0.42 & 0.12 & 29.01 & 8.86 & 2,928.88 & 128.75 \\
Standard deviation & 0.64 & 0.39 & 16.10 & 10.60 & 2,357.84 & 64.01 \\
Pearson correlation & \multicolumn{2}{c}{0.71} & \multicolumn{2}{c}{-0.86} & \multicolumn{2}{c}{0.68} \\
Spearman correlation & \multicolumn{2}{c}{0.77} & \multicolumn{2}{c}{-0.90} & \multicolumn{2}{c}{0.86} \\
\botrule
\end{tabular}
\end{adjustbox}
\end{center}
\end{table}

\subsection{Multivariate transformation}\label{subsec2.2}

\subsubsection{Rotation based iterative Gaussianisation}\label{subsubsec2.2.1}

RBIG is an iterative algorithm that applies marginal Gaussianisation followed by an orthonormal rotation. Although \cite{bib26} demonstrated that any orthonormal rotation, even a simple random rotation matrix, could be used in RBIG, PCA and ICA are more suitable. The choice of a rotation matrix is important and is usually based on multiple factors. PCA provides a suboptimal convergence rate compared to ICA and requires more iterations. However, convergence in ICA takes more time, especially in higher dimensional cases. In this study, we used RBIG with both PCA (RBIGP) and ICA (RBIGI) rotations for comparison. For RBIGP, the original MATLAB code from \cite{bib26} was implemented with histogram equalisation for marginal Gaussianisation. For RBIGI, a Python implementation was used with a normal score transformation for marginal Gaussianisation together with fast ICA \citep{bib32} from the Scikit-learn library \citep{bib33} for rotation.

The steps in RBIG are as follows:

\begin{enumerate}[1.]

\item First marginal Gaussianisation

\begin{equation}
Y_{(0)}=\Psi_{(0)}(Z),\label{eq1}
\end{equation}

where $\Psi_{(0)}(Z)$ is a normal score transformation for RBIGI or histogram equalisation for RBIGP applied to each dimension of the original data $Z$ before the first iteration.

\item ICA or PCA rotation expressed by an orthonormal rotation matrix $R_{(i)}$ at each iteration $i$

\begin{equation}
Y_{(i+1)}^{Rot}=R_{(i)}.\label{eq2}
\end{equation}

\item Marginal Gaussianisation of rotated variables at each iteration $i$

\begin{equation}
Y_{(i+1)}^{Gaus}=\Psi_{(i)}(Y_{(i+1)}^{Rot}).\label{eq3}
\end{equation}

\item Repeat steps 2 and 3 for a predefined number of iterations.

\end{enumerate}

In this study, the algorithm runs for a fixed number of iterations without any stopping criteria. In addition, RBIG saves rotation matrices and Gaussian tables at each iteration, including the first marginal Gaussianisation. Finally, the back transformation to the original state is performed in reverse order of the data saved at each iteration.

\subsubsection{Projection pursuit multivariate transform}\label{subsubsec2.2.2}

PPMT methodology is based on iteratively searching for interesting projections followed by normal score transformation along those projections \citep{bib13,bib16}. An interesting projection is one that has a maximum departure from Gaussianity based on the projection index developed by \cite{bib18}. The input parameters for the original algorithm are a maximum number of iterations and stopping criteria. The algorithm terminates after reaching the targeted projection index percentile based on the bootstrapping algorithm or after a set number of iterations \citep{bib13}. A detailed description and visualization of PPMT and its stopping criteria can be found in \cite{bib16}.

Similar to RBIG, PPMT runs for a fixed number of iterations in this study. The purpose of doing so is to check the multivariate normality of the transformed RBIG and PPMT factors after the same number of iterations. PPMT records rotation matrices and Gaussian tables at each step, which are used during the back transformation following the above steps in reverse order. We used a Python implementation of PPMT based on the original Fortran code from \cite{bib13}.

\subsubsection{Flow transformation}\label{subsubsec2.2.3}

The FA methodology differs from that of RBIG and PPMT, which are based on orthonormal rotations and marginal Gaussianisations. FA continuously deforms the original kernel density function into standard multi-Gaussian space using Lagrangian mechanics \citep{bib17}. The two main input parameters that characterise this deformation are the starting $\sigma_0$ and the final $\sigma_1$ spreads of the kernel that control the smoothing. This means that $\sigma_0$ controls how strongly FA deforms the kernels, for which a smaller value results in more Gaussian factors. In contrast, $\sigma_1$ controls the ranges of the produced factors, and it is recommended that $\sigma_1=\sigma_0+1$ so that the marginal distributions of the transformed data have standard deviations close to 1 \citep{bib35}. For the details of the methodology of the FA algorithm, readers are referred to the original literature \citep{bib17,bib36}.

In this study, the ``gmGeostats” CRAN package \citep{bib36} was used for FA and the transformation was chained twice to achieve multivariate normality.

\subsubsection{Chained multi-Gaussian transform and spatial decorrelation}\label{subsubsec2.2.4}

MGT methods, similar to PCA and ICA linear transforms, can only guarantee a decorrelation at lag zero. A practical solution to ensure spatial decorrelation is to chain those methods with MAF \citep{bib10}. The original geostatistical application of MAF uses PCA twice, once on a covariance matrix at lag 0 and once on a single non-zero lag of the cross-variogram function. However, as RBIG, PPMT or FA already produce independent factors, only the second MAF is used. Chained MGT and MAF comprise the following steps:

\begin{enumerate}[1.]

    \item Transform the original data $Z$ into independent standard multi-Gaussian variables $Y^{MG}$.
    
    \item Calculate a sphering matrix $S^{-1/2}$ for multi-Gaussian factors:

    \begin{equation} 
    S^{-1/2}=Q\Lambda^{-1⁄2}Q^T,\label{eq4}
    \end{equation}

    where $Q$ is an eigenvector matrix and $\Lambda$ is the corresponding diagonal matrix.
    
    \item Compute an eigenvector matrix $Q_h$ from the spectral decomposition of the cross-variogram matrix at lag $h$.
    
    \item Multiply the multi-Gaussian factors by the sphering and eigenvector matrices
    
    \begin{equation} 
    Y^{MAF}=S^{-1/2}Q_hY^{MG}.\label{eq5}
    \end{equation}
    
\end{enumerate}

The resulting factors will also be spatially independent. Back transformation is performed by using transposed sphering and eigenvector matrices. A combination of MAF with MGT has been applied in various geostatistical case studies \citep{bib21,bib25,bib36}. However, MAF is applied on a single non-zero lag distance, so its spatial decorrelation is not perfect at all lags.

\subsection{Metrics for comparison}\label{subsec2.3}

In this study, RBIGP, RBIGI, PPMT and FA were compared using six metrics: multivariate normality, spatial decorrelation, execution times, qualitative assessment of the reproduction of multivariate distributions, and histogram and variogram validation. Multivariate normality (MVN) can be evaluated using different multivariate normality tests from the MVN R package \citep{bib37}. The MVN tests in this package include Mardia’s measures of multivariate skewness and kurtosis \citep{bib38}, Royston’s techniques for assessing multivariate normality \citep{bib39}, the Henze-Zirkler invariant consistent tests for multivariate normality \citep{bib40} and Energy statistics \citep{bib41}. Many studies in geostatistics have used MVN tests to assess the performance of MGT e,g., \citep{bib17,bib19,bib36}. In this study, only Henze-Zirkler’s and Energy tests were used based on their consistency and robustness, confirmed in some comparison and review articles \citep{bib43,bib42}.

A common problem with MGT methods is that they cannot ensure spatial decorrelation. In a geostatistical context, spatial decorrelation is sometimes more crucial than multivariate normality. It is, therefore, important to assess spatial decorrelation even after applying MAF. For this purpose, the quality of the spatial decorrelation was evaluated through experimental cross-variograms. In addition, a quantitative assessment of spatial decorrelation was conducted using the relative deviation from diagonality $\tau(h)$ together with the spatial diagonalisation efficiency $\kappa(h)$ measures suggested by \cite{bib51}. The first measure compares the absolute sum of off-diagonal elements in the factor variogram matrix to the corresponding diagonal elements. The second one compares the sum of squares of off-diagonal elements in the factor variogram matrix to the sum of squares of off-diagonal elements in the sample variogram matrix. Perfect spatial decorrelation will result in zero for $\tau(h)$ and one for $\kappa(h)$.

The execution time for both forward and inverse multi-Gaussian transformations is another important factor. RBIGI, PPMT and FA were applied in a Jupyter Notebook environment, where their CPU times could be recorded. FA in the gmGeostats R package \citep{bib36} and the Python implementations of RBIGI and PPMT were used for the work reported in this paper. Although the original Fortran program for PPMT is much faster, the Python code may provide a fairer comparison of PPMT and RBIGI. For RBIGP, the original MATLAB code \citep{bib26} was used, in which CPU time can also be measured.

Finally, reproduction of histograms and variograms are the most critical geostatistical properties. Even though their reproduction can be assessed qualitatively, the root mean square error (RMSE) was used to obtain a quantitative comparison of results. For example, the RMSE between a hundred percentiles of original and simulated cumulative distribution functions (CDFs) were calculated for histogram validation. Similarly, RMSE measures were calculated between experimental and simulated direct and cross-variograms at multiple lag distances. Metrics such as these can be found in other geostatistical studies \citep{bib11,bib25}. However, it is more difficult to provide quantitative comparisons of the reproduction of multivariate complexities, which is the primary objective of this study. For this purpose, cross-plots between simulated variables were qualitatively assessed and compared with the original plots from Figure \ref{fig1}.

\section{Results}\label{sec3}

\subsection{Multi-Gaussian transformation}\label{subsec3.1}

\begin{figure}[ht]%
\centering
\includegraphics[width=1\textwidth]{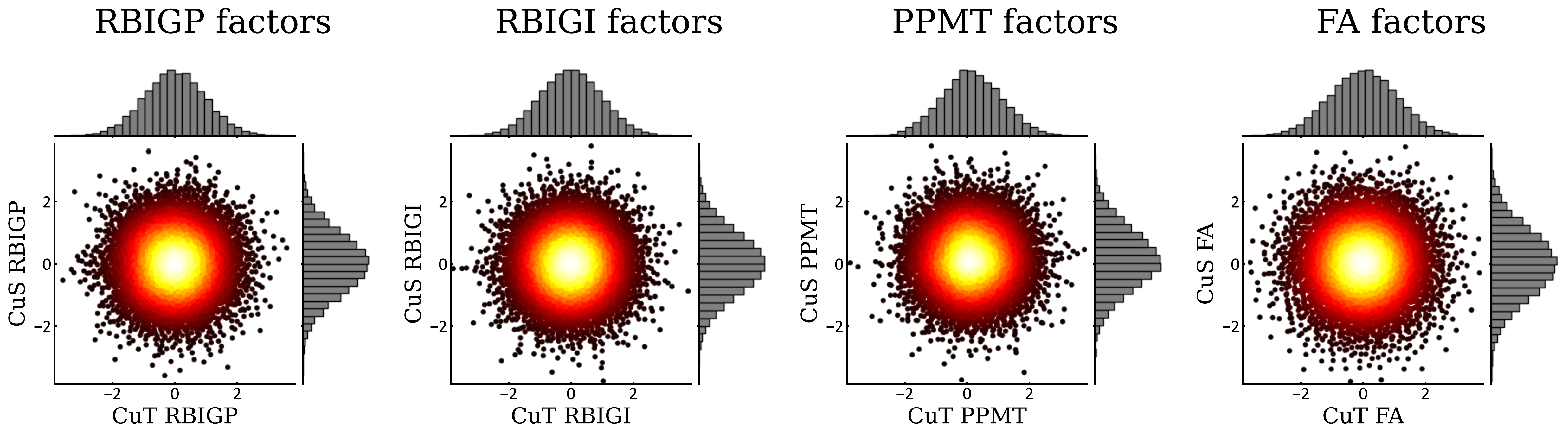}
\caption{Cross plots of multi-Gaussian transformation results for case A.}\label{fig2}
\end{figure}

RBIGP, RBIGI, PPMT and FA were applied to three case studies with multivariate complexities. 150 iterations were used for RBIGP, RBIGI and PPMT. For FA, input parameters were $\sigma_0=0.1$ and $\sigma_1=1.1$ and this method was chained twice to achieve a standard multi-Gaussian distribution. As a result, all transforms produced independent multi-Gaussian factors for total and soluble copper grades in case A (Figure \ref{fig2}). RBIGI required 34 seconds, while PPMT and FA required 24 and 18 seconds, respectively.

\begin{figure}[ht]%
\centering
\includegraphics[width=1\textwidth]{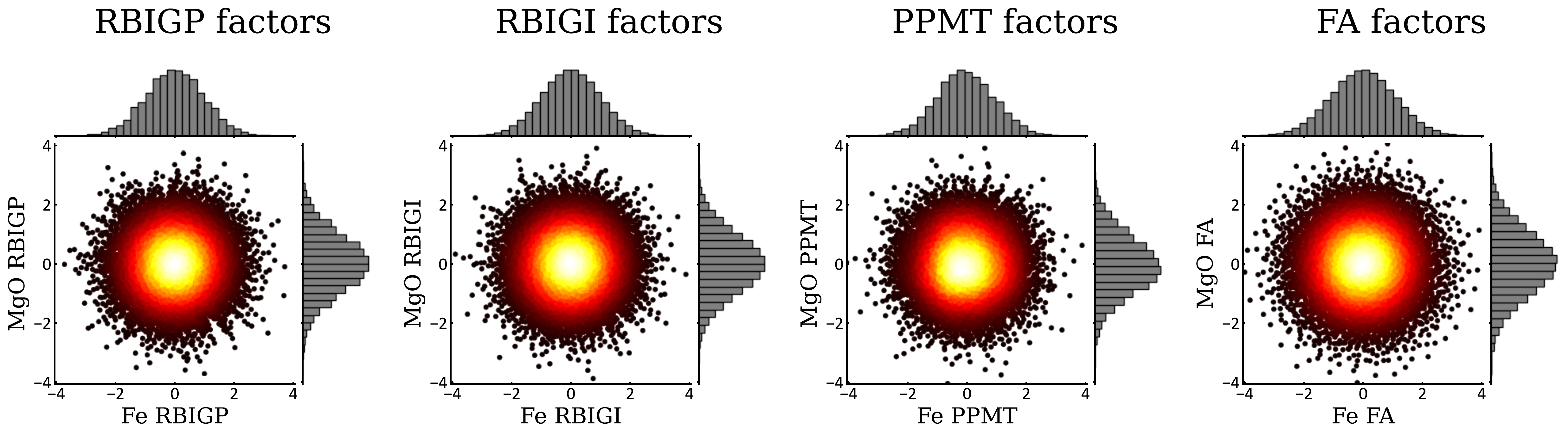}
\caption{Cross plots of multi-Gaussian transformation results for case B.}\label{fig3}
\end{figure}

In case B, the non-linear relationship between iron and magnesium oxide was transformed into a multi-Gaussian distribution by all three methods (Figure \ref{fig3}). However, because this case has 9,990 sample points, it took slightly more time to execute: 50 seconds for RBIGI, 48 seconds for PPMT and 51 seconds for FA.

\begin{figure}[ht]%
\centering
\includegraphics[width=1\textwidth]{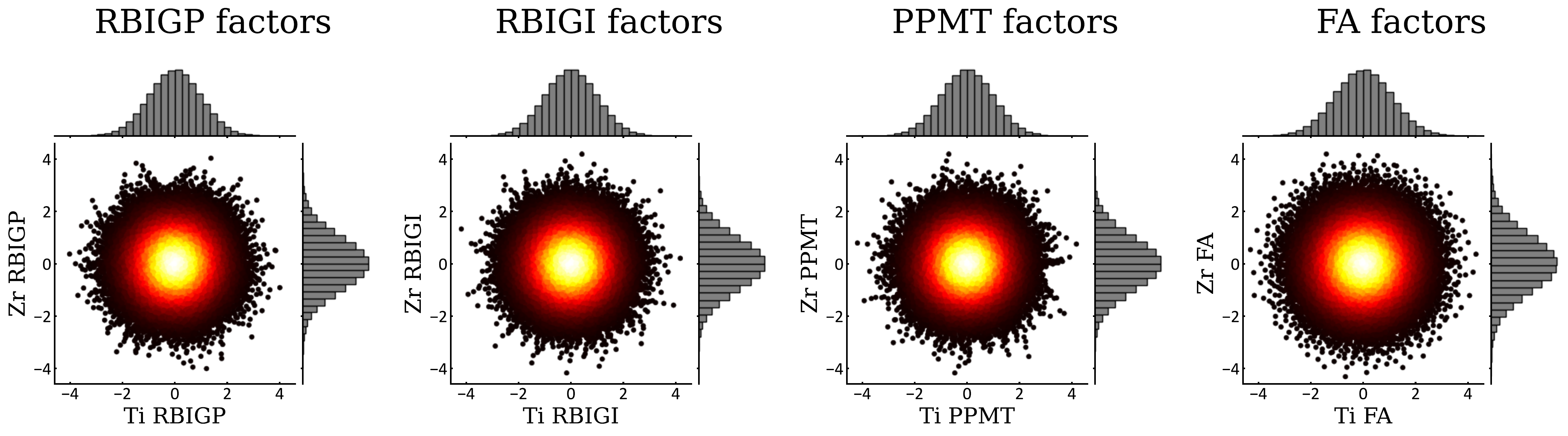}
\caption{Cross plots of multi-Gaussian transformation results for case C.}\label{fig4}
\end{figure}

Case C was more challenging to transform and Figure \ref{fig4} shows the transformed factors from all three methods. Declustering weights were not used in this case because the sum of the weights was not equal to the number of points. Furthermore, as case C comprises 33,021 titanium and zirconium samples, it takes more time to transform. For example, RBIGI and PPMT required 2 minutes 46 seconds and 2 minutes 42 seconds, respectively. On the other hand, it took FA 9 minutes 11 seconds to transform the heteroscedasticity between variables into independent factors with a multi-Gaussian distribution. In all three cases, RBIGP completed the transformation in less than one second. The significant difference in time between RBIGP and RBIGI can be explained by the different programming languages and simpler marginal Gaussianisation.

Visual inspection of kernel density estimates can assist in assessing the quality of results. However, a better way to compare the results is to check for multivariate normality. Table \ref{tab2} shows the results from MVN tests applied to the transformed factors. For cases A and C, all three transforms produced multi-Gaussian distributions with perfect p-values according to the Henze-Zirkler and Energy tests. PPMT, however, could not produce multi-Gaussian factors and failed both MVN tests (i.e., with p-values of less than 0.05). RBIGI very slightly outperformed the other transforms based on multi-Gaussianity, particularly in cases A and B. PPMT failed to produce an MVN distribution in case B and showed slightly worse results in case A. On the other hand, all three methods produced similar results for case C. It is also important to note that the FA factors do not have a unit covariance matrix and have a standard deviation of 1.09 in all three cases.

\begin{table}[h]
\begin{center}
\caption{Multivariate normality test results for three case studies}\label{tab2}%
\begin{adjustbox}{width=0.9\textwidth}
\begin{tabular}{@{}lccccc@{}}
\toprule
MVN test & Case & RBIGP & RBIGI & PPMT & FA \\
\midrule
\multirow{3}{*}{Henze-Zirkler’s} & A & 0.0932 (p = 1) & 0.0205 (p = 1) & 0.3023 (p = 1) & 0.0237 (p = 1) \\
& B & 0.0621 (p = 1) & 0.0226 (p = 1) & 1.2135 (p $<$ 0.05) & 0.0270 (p = 1) \\
& C & 0.3951 (p = 1) & 0.3362 (p = 1) & 0.3319 (p = 1) & 0.3198 (p = 1) \\
\multirow{3}{*}{Energy} & A & 0.0969 (p = 1) & 0.0394 (p = 1) & 0.2382 (p = 1) & 0.0860 (p = 1) \\
& B & 0.0845 (p = 1) & 0.0337 (p = 1) & 1.2746 (p $<$ 0.05) & 0.0740 (p = 1) \\
& C & 0.2835 (p = 1) & 0.2214 (p = 1) & 0.2307 (p = 1) & 0.2377 (p = 1) \\
\botrule
\end{tabular}
\end{adjustbox}
\end{center}
\end{table}

\subsection{Spatial decorrelation}\label{subsec3.2}

\begin{figure}[ht]%
\centering
\includegraphics[width=1\textwidth]{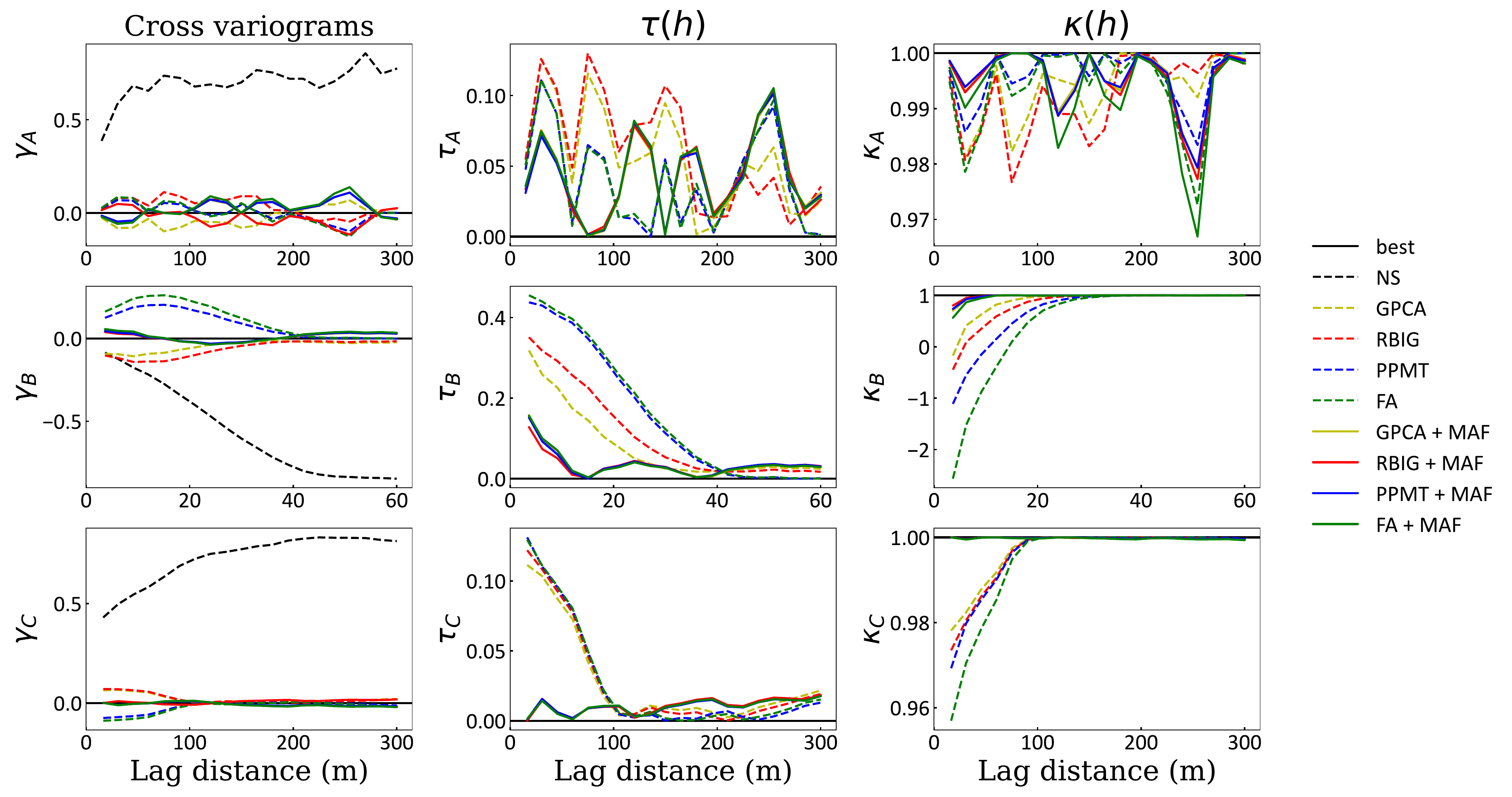}
\caption{Cross-variograms, relative deviation from diagonality $\tau(h)$ and spatial diagonalisation efficiency $\kappa(h)$ for case A (top row), case B (middle row) and case C (bottom row) before and after applying MAF.}\label{fig5}
\end{figure}

One of the limitations of the MGT approaches is that they can only assume spatial decorrelation. MGT methods always guarantee the decorrelation of variables at lag zero but not at further lags. Figure \ref{fig5} (left column) shows that the omni-directional cross-variograms of the MGT factors deviate slightly from zero in all three case studies. A better spatial decorrelation was achieved by chaining RBIGP, RBIGI, PPMT and FA with MAF. A 75 m lag for case A and a 15 m lag for cases B and C were selected for the MAF transformation, and the resulting matrices are shown in Table \ref{tab3}. As a result, the decorrelated factors for cases B and C are more spatially independent, whereas case A did not show much improvement. Nevertheless, the resulting factors are significantly more decorrelated than the normal score variograms.

\begin{table}[ht]
\begin{center}
\caption{Computed MAF matrices for multi-Gaussian factors in the three case studies}\label{tab3}%
\begin{adjustbox}{width=0.8\textwidth}
\begin{tabular}{@{}lccccc@{}}
\toprule
Case & Lag & RBIGP & RBIGI & PPMT & FA \\
\midrule
A & 75 & $\begin{bmatrix}-0.90&-0.45\\0.45&-0.90\\\end{bmatrix}$ & $\begin{bmatrix}-0.86&0.53\\-0.53&-0.86\\\end{bmatrix}$ & $\begin{bmatrix}0.24&-0.98\\0.98&0.24\\\end{bmatrix}$ & $\begin{bmatrix}0.23&-0.98\\0.98&0.23\\\end{bmatrix}$ \\
\addlinespace
B & 15 & $\begin{bmatrix}-0.20&-0.99\\0.99&-0.20\\\end{bmatrix}$ & $\begin{bmatrix}-0.34&-0.95\\0.95&-0.34\\\end{bmatrix}$ & $\begin{bmatrix}-0.79&0.64\\-0.64&-0.79\\\end{bmatrix}$ & $\begin{bmatrix}-0.78&0.65\\-0.65&-0.78\\\end{bmatrix}$ \\
\addlinespace
C & 15 & $\begin{bmatrix}0.47&-0.89\\0.89&0.47\\\end{bmatrix}$ & $\begin{bmatrix}-0.84&0.55\\-0.55&-0.84\\\end{bmatrix}$ & $\begin{bmatrix}-0.77&-0.66\\0.66&-0.77\\\end{bmatrix}$ & $\begin{bmatrix}-0.78&-0.65\\0.65&-0.78\\\end{bmatrix}$ \\
\botrule
\end{tabular}
\end{adjustbox}
\end{center}
\end{table}

Another way to assess spatial decorrelation is by checking the relative deviation from diagonality $\tau(h)$ and spatial diagonalisation efficiency $\kappa(h)$ introduced by \cite{bib51}. Figure \ref{fig5} (middle and right columns) shows that chaining MGT with MAF improves spatial decorrelation. Even though case A does not appear to be much better, there is a clear decorrelation at lower lags up to 100 m after applying MAF. It should also be noted that MGT methods alone are not sufficient to ensure decorrelation in case B. This is evident by $\tau(h)$ and $\kappa(h)$ not being close to zero and one, respectively. Such poor results can be explained by the normal score cross-variograms being very close to the factor cross-variograms. The average $\tau(h)$ and $\kappa(h)$ results are shown in Table \ref{tab4}.

\begin{table}[ht]
\begin{center}
\caption{Average relative deviation from diagonality $\tau(h)$ and spatial diagonalisation efficiency $\kappa(h)$}\label{tab4}%
\begin{adjustbox}{width=0.7\textwidth}
\begin{tabular}{@{}lcccccc@{}}
\toprule
\multirow{2}{*}{Method} & \multicolumn{2}{c}{Case A} & \multicolumn{2}{c}{Case B} & \multicolumn{2}{c}{Case C} \\
& $\overline{\tau}$ & $\overline{\kappa}$ & $\overline{\tau}$ & $\overline{\kappa}$ & $\overline{\tau}$ & $\overline{\kappa}$ \\
\midrule
RBIGP & 0.0550 & 0.9939 & 0.0825 & 0.8755 & 0.0285 & 0.9967 \\
RBIGI & 0.0606 & 0.9924 & 0.1106 & 0.8046 & 0.0287 & 0.9963 \\
PPMT & 0.0394 & 0.9960 & 0.1595 & 0.6562 & 0.0275 & 0.9960 \\
FA & 0.0397 & 0.9940 & 0.1662 & 0.4299 & 0.0279 & 0.9942 \\
RBIGP+MAF & 0.0416 & 0.9956 & 0.0364 & 0.9780 & 0.0105 & 0.9998 \\
RBIGI+MAF & 0.0424 & 0.9952 & 0.0338 & 0.9852 & 0.0107 & 0.9998 \\
PPMT+MAF & 0.0418 & 0.9956 & 0.0363 & 0.9805 & 0.0103 & 0.9998 \\
FA+MAF & 0.0426 & 0.9933 & 0.0364 & 0.9668 & 0.0103 & 0.9997 \\
\botrule
\end{tabular}
\end{adjustbox}
\end{center}
\end{table}

\subsection{Geostatistical conditional simulation}\label{subsec3.3}

\begin{figure}[ht]%
\centering
\includegraphics[width=0.8\textwidth]{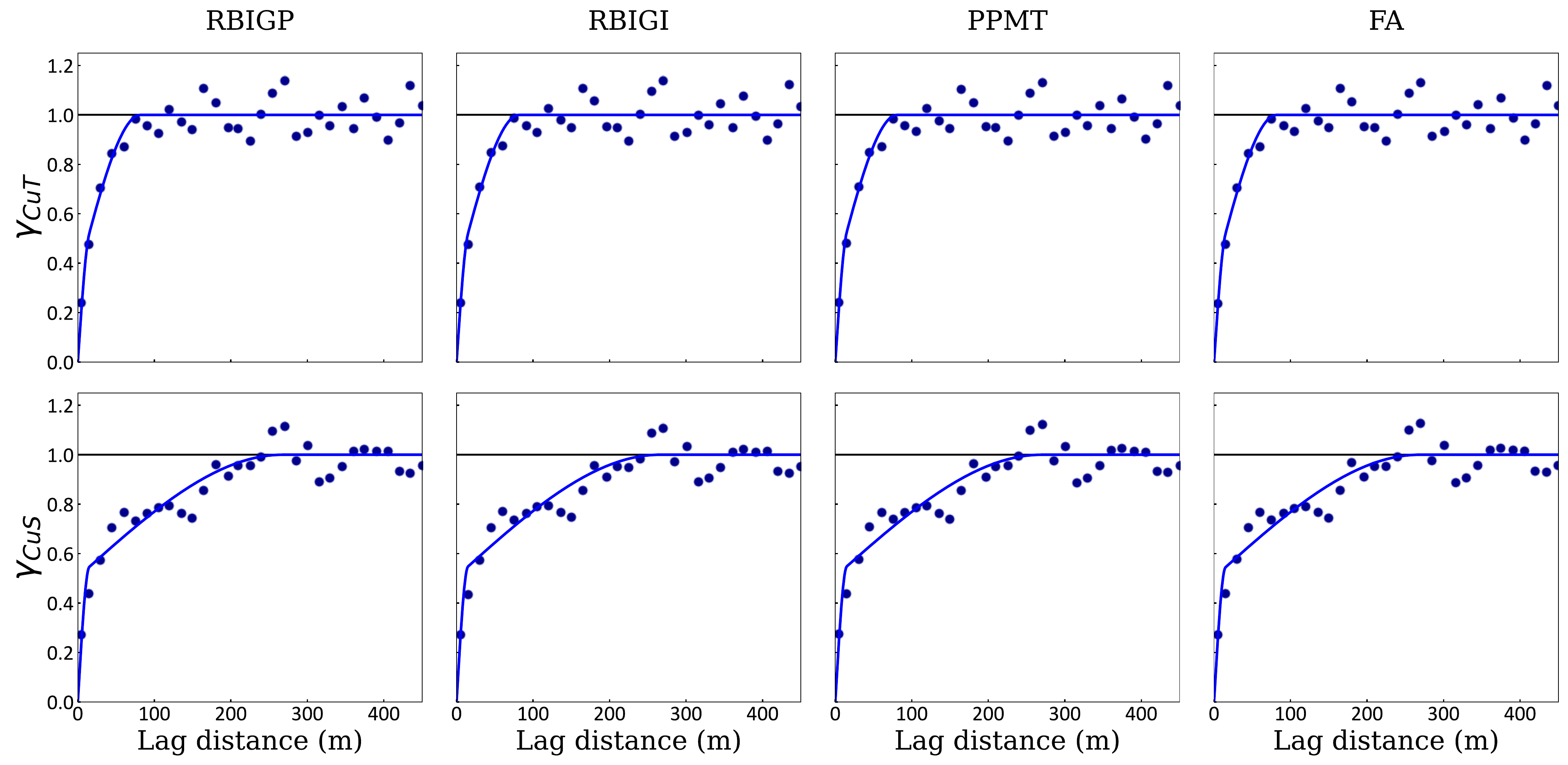}
\caption{Direct variograms of decorrelated factors of total (top row) and soluble (bottom row) copper grades for case A. Experimental variograms are indicated with points and fitted theoretical variograms with solid lines.}\label{fig6}
\end{figure}

Direct variograms were automatically fitted before conditionally simulating the decorrelated variables. Various automated and semi-automated variogram fitting algorithms can be used for this purpose \citep{bib44,bib45} and are available in various commercial software packages. There was no significant directional anisotropy for case A, and omni-directional variograms were modelled for the RBIG, PPMT and FA factors (Figure \ref{fig6}). The spatial variability is very similar among the three transforms, which means that any differences between generated realisations will be due mainly to the back-transformations.

\begin{equation}
\begin{aligned}
\gamma_{{CuT}_{RBIGP}}=0.342Sph(15m)+0.658Sph(82m) \\
\gamma_{{CuS}_{RBIGP}}=0.505Sph(15m)+0.495Sph(267m) \\
\gamma_{{CuT}_{RBIGI}}=0.342Sph(15m)+0.658Sph(82m) \\
\gamma_{{CuS}_{RBIGI}}=0.506Sph(15m)+0.494Sph(265m) \\
\gamma_{{CuT}_{PPMT}}=0.347Sph(15m)+0.653Sph(81m) \\
\gamma_{{CuS}_{PPMT}}=0.507Sph(15m)+0.493Sph(269m) \\
\gamma_{{CuT}_{FA}}=0.340Sph(15m)+0.660Sph(82m) \\
\gamma_{{CuS}_{FA}}=0.503Sph(15m)+0.497Sph(268m)
\end{aligned}
\label{eq6}
\end{equation}

\begin{figure}[ht]%
\centering
\includegraphics[width=0.8\textwidth]{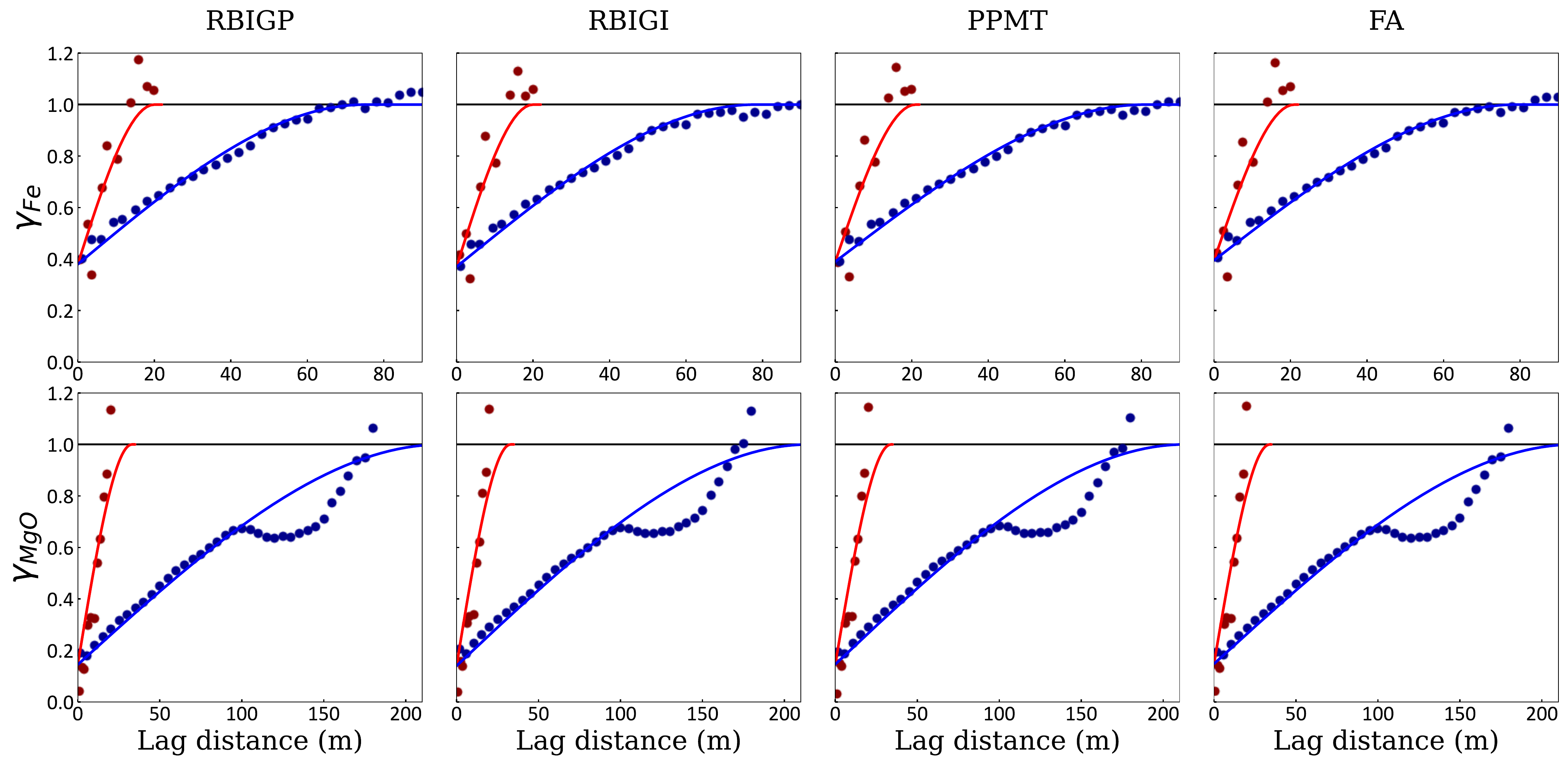}
\caption{Direct variograms of decorrelated factors for iron (top row) and magnesium oxide (bottom row) for case B. Experimental variograms are indicated by points and fitted theoretical variograms by solid lines (blue: horizontal, and red: vertical).}\label{fig7}
\end{figure}

On the other hand, the data in case B showed significantly different variabilities in horizontal and vertical directions. Experimental variograms of decorrelated iron and magnesium oxide factors were automatically fitted as shown in Figure \ref{fig7}. However, for MAF to provide a decorrelation at all lags, spatial variability must be represented by a two-structured LMC \citep{bib10}. In cases A and B, the co-regionalisation models appear to be more complex and thus experimental variograms could not be fitted to two nested structures. Nevertheless, chaining MAF and MGT still demonstrates better results, even when two-structured LMC cannot be produced \citep{bib21,bib25}.

\begin{equation}
\begin{aligned}
\gamma_{{Fe}_{RBIGP}}=0.380Nug+0.620Sph(20m,75m) \\
\gamma_{{MgO}_{RBIGP}}=0.145Nug+0.855Sph(33m,222m) \\
\gamma_{{Fe}_{RBIGI}}=0.370Nug+0.630Sph(20m,78m) \\
\gamma_{{MgO}_{RBIGI}}=0.140Nug+0.860Sph(33m,215m) \\
\gamma_{{Fe}_{PPMT}}=0.390Nug+0.610Sph(21m,81m) \\
\gamma_{{MgO}_{PPMT}}=0.146Nug+0.854Sph(34,213m) \\
\gamma_{{Fe}_{FA}}=0.395Nug+0.605Sph(21m,79m) \\
\gamma_{{MgO}_{FA}}=0.148Nug+0.852Sph(34m,220m)
\end{aligned}
\label{eq7}
\end{equation}

\begin{figure}[ht]%
\centering
\includegraphics[width=0.8\textwidth]{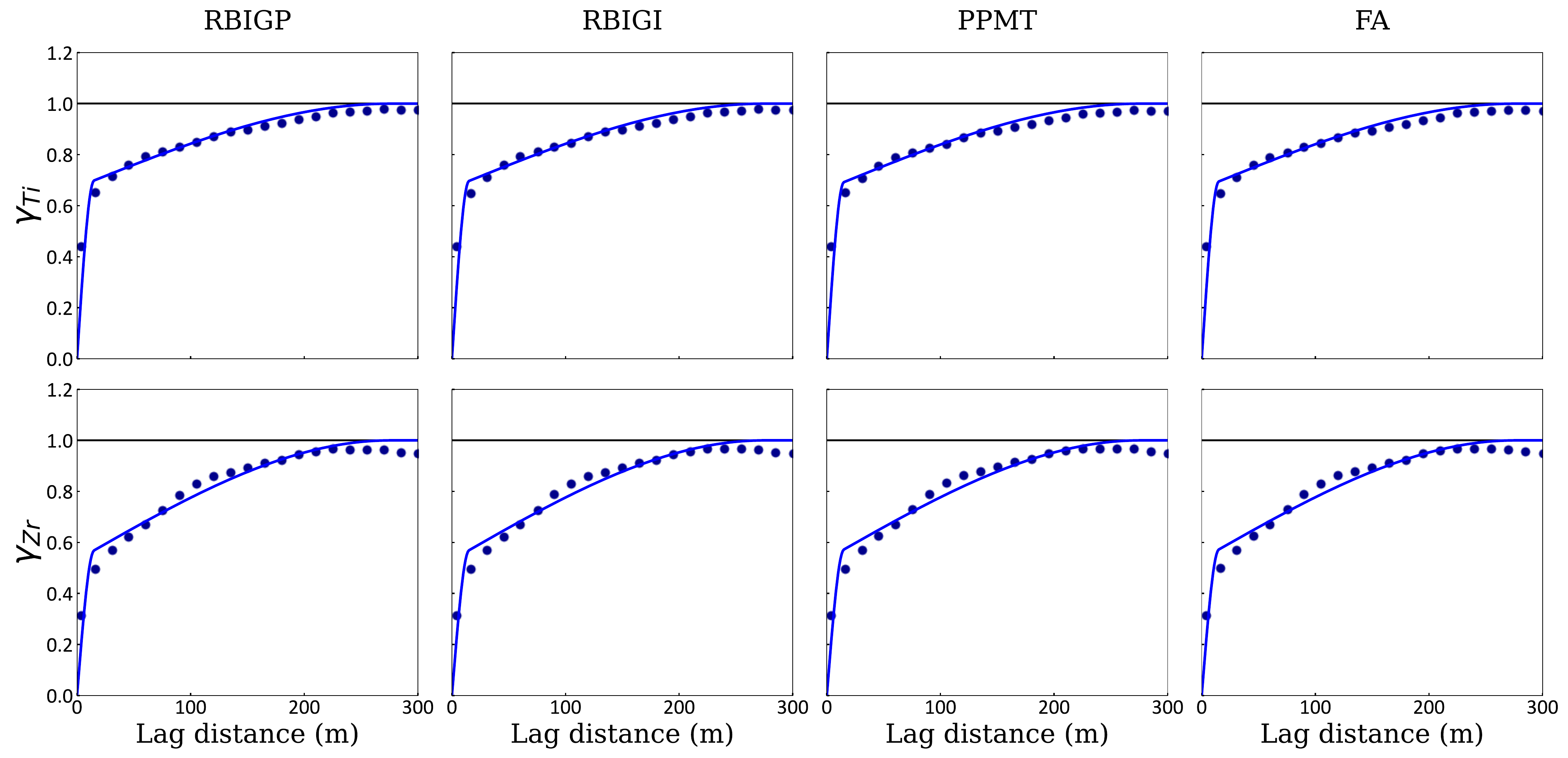}
\caption{Direct variograms of decorrelated factors of titanium (top row) and zirconium (bottom row) for case C. Experimental variograms are indicated by points and the fitted theoretical variograms by solid lines.}\label{fig8}
\end{figure}

Finally, omni-directional experimental variograms were calculated for case C. Figure \ref{fig8} shows the variograms for the RBIG, PPMT and FA factors. Unlike the other two cases, variograms in case C were fitted to two nested structures.

\begin{equation}
\begin{aligned}
\gamma_{{Ti}_{RBIGP}}=0.672Sph(15m)+0.328Sph(275m) \\
\gamma_{{Zr}_{RBIGP}}=0.530Sph(15m)+0.470Sph(275m) \\
\gamma_{{Ti}_{RBIGI}}=0.670Sph(15m)+0.330Sph(274m) \\
\gamma_{{Zr}_{RBIGI}}=0.531Sph(15m)+0.469Sph(274m) \\
\gamma_{{Ti}_{PPMT}}=0.665Sph(15m)+0.335Sph(275m) \\
\gamma_{{Zr}_{PPMT}}=0.534Sph(15m)+0.466Sph(275m) \\
\gamma_{{Ti}_{FA}}=0.668Sph(15m)+0.332Sph(275m) \\
\gamma_{{Zr}_{FA}}=0.534Sph(15m)+0.466Sph(275m)
\end{aligned}
\label{eq8}
\end{equation}

It is evident that the factors produced by each method have almost identical spatial properties, which is true for all three cases. There were only minor differences in the variogram parameters generated by automated fitting. Using these variogram models, conditional turning bands simulation \citep{bib46,bib47} was used to generate 50 realisations. In each case, the produced multi-Gaussian factors were modelled using identical neighbourhood parameters. Thus, most of the differences between realisations will be due to the variable transformation, which ensures a fair comparison of the MGT methods.

For case A, simulations were run on a 61 × 115 × 38 block model with grid dimension of 30m × 30m × 10m. A moving neighbourhood of 1000m × 1000m × 1000m with 8 octants and 8 points per octant was used for turning bands simulation. The block model in case B consists of 64 × 69 × 39 grids with 3m × 3m × 2m size, for which a 200m × 200m × 200m moving neighbourhood with 8 octants and 16 points per octant was used. Finally, realisations for case C were produced on a 65 × 36 × 54 block model with a 15m × 15m × 15m grid size. A 450m × 450m × 450m moving neighbourhood was used with 8 octants and 40 points per octant.

\subsection{Analysis and validation}\label{subsec3.4}

The simulated realisations were back-transformed to the original scale. To do so, MGT+MAF factors were first multiplied by the transposed MAF matrices (see Table \ref{tab3}), Then the RBIGP, RBIGI, PPMT and FA inverse transformations were applied to the corresponding multi-Gaussian realisations. In case A, RBIGP and PPMT required only 3 seconds to back-transform a single realisation and RBIGI required 22 seconds, whereas FA took 13 minutes and 29 seconds. A similar difference was observed in case B, where FA required 14 minutes and 30 seconds, RBIGI required 13 seconds and both RBIGP and PPMT required 2 seconds to back-transform one realisation. Case C is much larger than the other two, but RBIGP, RBIGI and PPMT needed only 1, 7 and 3 seconds, respectively. This is because their back-transformation does not require the original data, only Gaussian tables and rotation matrices at each iteration. However, as the original data are used in the FA back transformation, it took 34 minutes and 59 seconds to back-transform a single realisation for case C.    

\subsubsection{Reproduction of bivariate relationships}\label{subsubsec3.4.1}

\begin{figure}[ht]%
\centering
\includegraphics[width=1\textwidth]{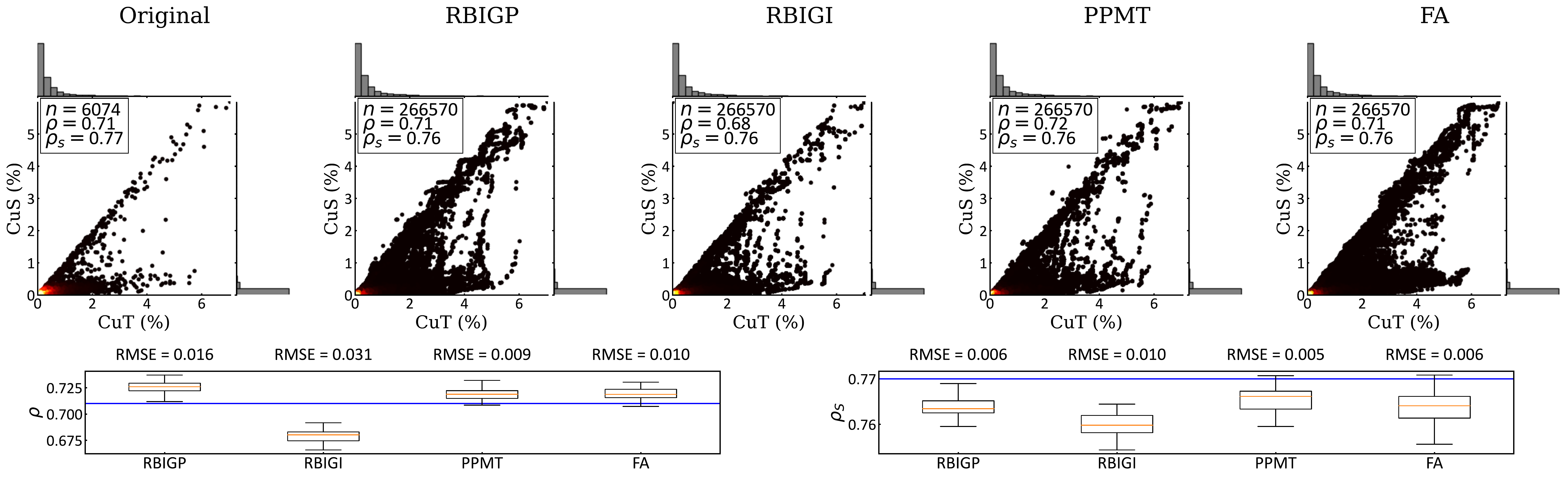}
\caption{Top: cross plots of back-transformed simulated grades from one realisation compared to the original data in case A. Bottom: box plots of Pearson $\rho$ and Spearman $\rho_S$ correlation coefficients from 50 realisations of case A compared to the original values (solid blue lines).}\label{fig9}
\end{figure}

Figure \ref{fig9} (top) shows the cross plots of the back-transformed variables from a single realisation of case A. The bivariate relationships of the simulated data are similar to those of the original distribution. Although there are some artefacts above an inequality constraint, this is not surprising given the skewed distributions of the copper grades. In fact, even acceptance-rejection methods that reject and re-simulate values to be within the bounds of inequality constraints do so at the cost of other statistical properties. For example, the acceptance-rejection approach performs well when marginal distributions are moderately skewed \citep{bib29}, but poorly reproduces other properties when dealing with very skewed distributions \citep{bib30}. Finally, reproduction of the Pearson and Spearman correlation coefficients suggests that PPMT performed slightly better than the others, while RBIGI underestimated the correlations (Figure \ref{fig9} bottom).

\begin{figure}[ht]%
\centering
\includegraphics[width=1\textwidth]{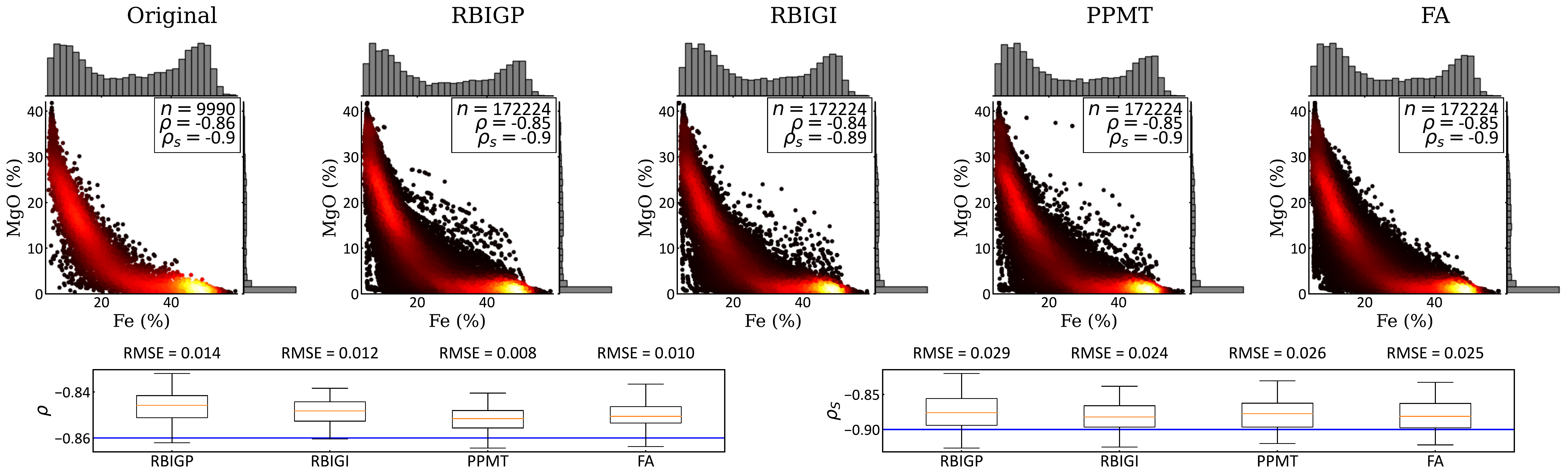}
\caption{Top: cross plots of back-transformed simulated grades from one realisation compared to the original data in case B. Bottom: box plots of Pearson $\rho$ and Spearman $\rho_S$ correlation coefficients from 50 realisations of case B compared to the original values (solid blue lines).}\label{fig10}
\end{figure}

In case B, the back-transformed results show a non-linear bivariate relationship, similar to the original data (Figure \ref{fig10}). Moreover, the reproduction of correlation coefficients is also almost identical.

\begin{figure}[ht]%
\centering
\includegraphics[width=1\textwidth]{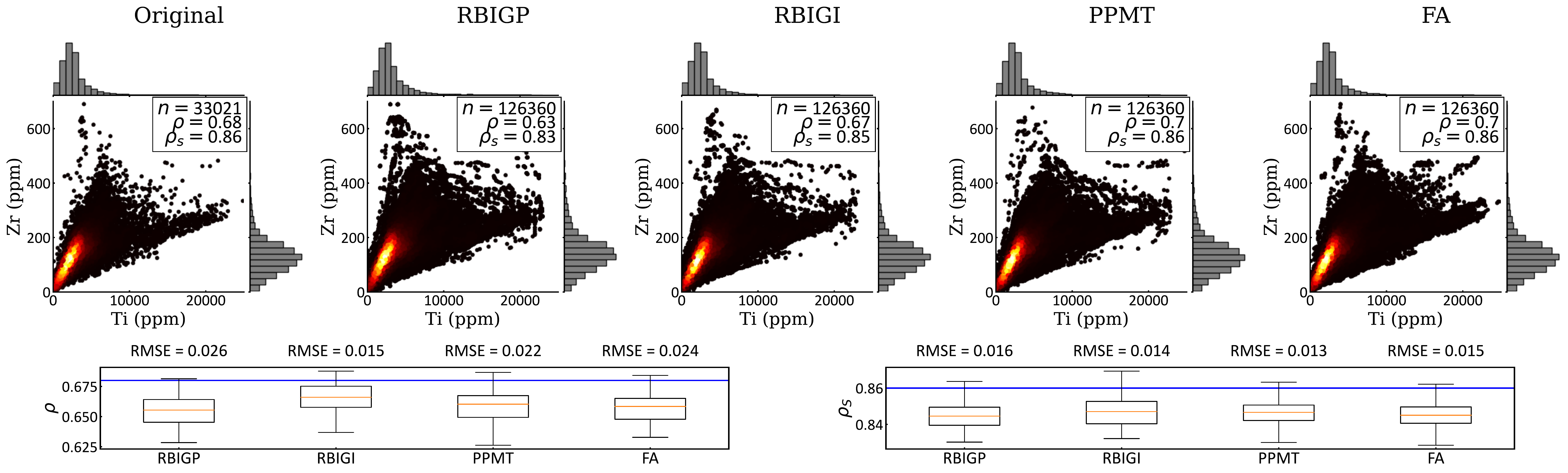}
\caption{Top: cross plots of back-transformed simulated grades from one realisation compared to the original data in case C. Bottom: box plots of Pearson $\rho$ and Spearman $\rho_S$ correlation coefficients from 50 realisations of case B compared to the original values (solid blue lines).}\label{fig11}
\end{figure}

Finally, the cross plots in Figure \ref{fig11} demonstrate that all the presented methods reproduced the heteroscedastic relationship between titanium and zirconium in case C. It can also be observed in the reproduction of the Pearson and Spearman correlation coefficients, where all three methods show similar results. However, despite similar correlations, a visual inspection suggests that FA produces better results. RBIGI produced significant outliers in the top left and bottom right corners of its cross plots in cases A and the top right corner in case C (Figures \ref{fig9} and \ref{fig11}). On the other hand, RBIGP and PPMT produced minor artefacts in all three cases, which was not observed in the FA realisations.

\subsubsection{Histogram and variogram validation}\label{subsubsec3.4.2}

\begin{figure}[ht]%
\centering
\includegraphics[width=1\textwidth]{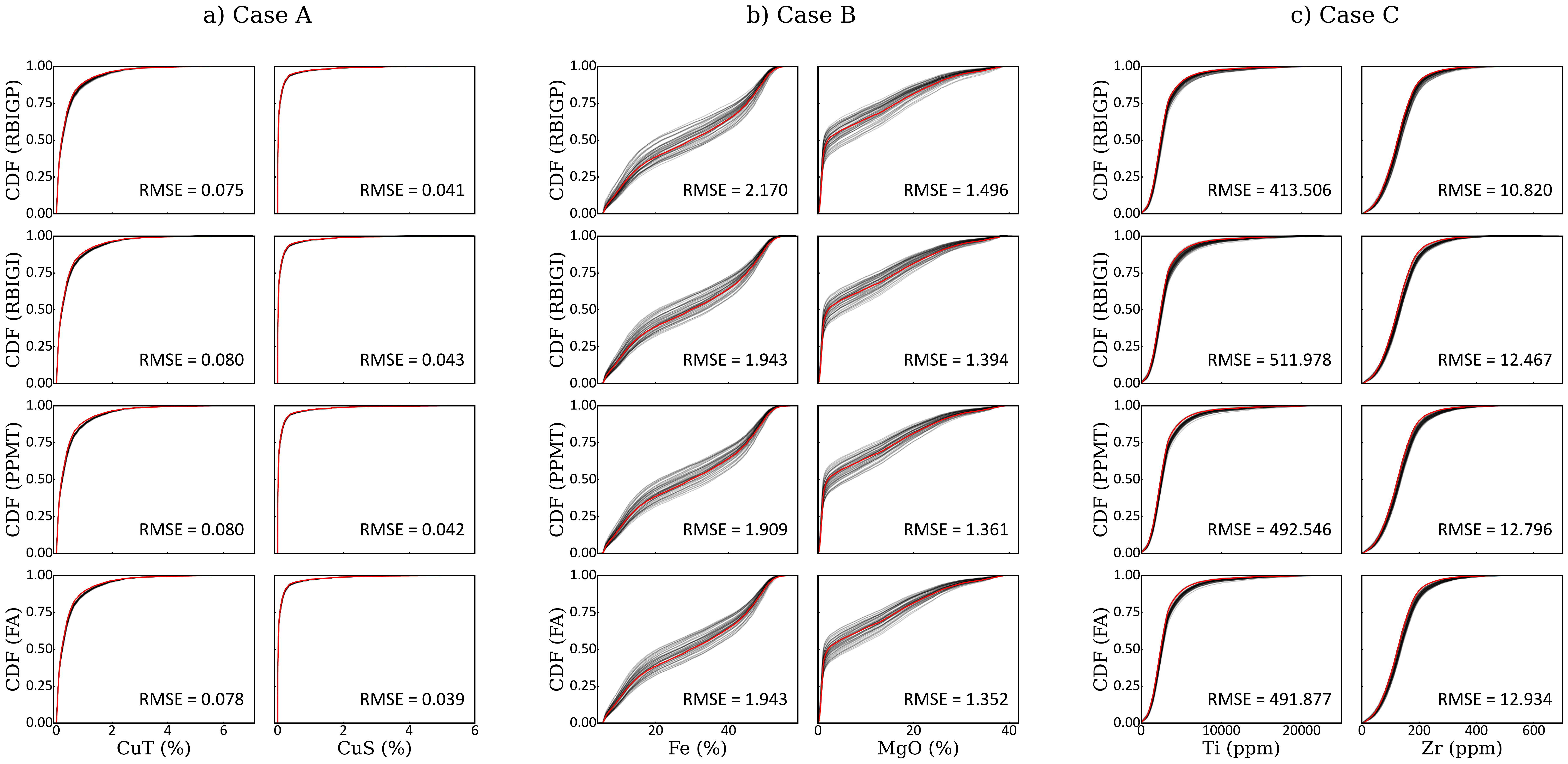}
\caption{Histogram validation for case A (a), case B (b) and case C (c). The CDFs of the simulated grades are shown as solid black lines and the solid red lines are the original functions.}\label{fig12}
\end{figure}

Histogram reproduction is another important part of geostatistical validation. Histograms of back-transformed variables of a single realization were well reproduced in all three case studies (see Figures \ref{fig9}, \ref{fig10} and \ref{fig11}). However, it is more appropriate to check all realisations, which can be accomplished by CDF or Q-Q plots. Figure \ref{fig12} shows the reproduction of the CDFs of the back-transformed realisations and the RMSE values between percentiles of the realisations and the original distributions. In case A, the skewness of the distributions makes it challenging to assess and compare, but RMSE results suggest that RBIGP and FA performed slightly better than the others. Similarly, RBIGP has the least RMSE for both variables in case C, but other methods performed better in case B. Nevertheless, the difference in RMSE is insignificant, and the overall results are similar.

Reproduction of spatial variability was checked on the normal scores (i.e., first normal score transformation before iterations of RBIGP, RBIGI, PPMT and FA). The reason for this is to avoid significant deviations due to the skewness of the original variables. In addition, experimental normal score variograms were used as theoretical variograms of normal scores were not modelled. Finally, RMSE between variogram values for simulated and experimental variograms were calculated to provide a quantitative comparison of result. 

\begin{figure}[ht]%
\centering
\includegraphics[width=1\textwidth]{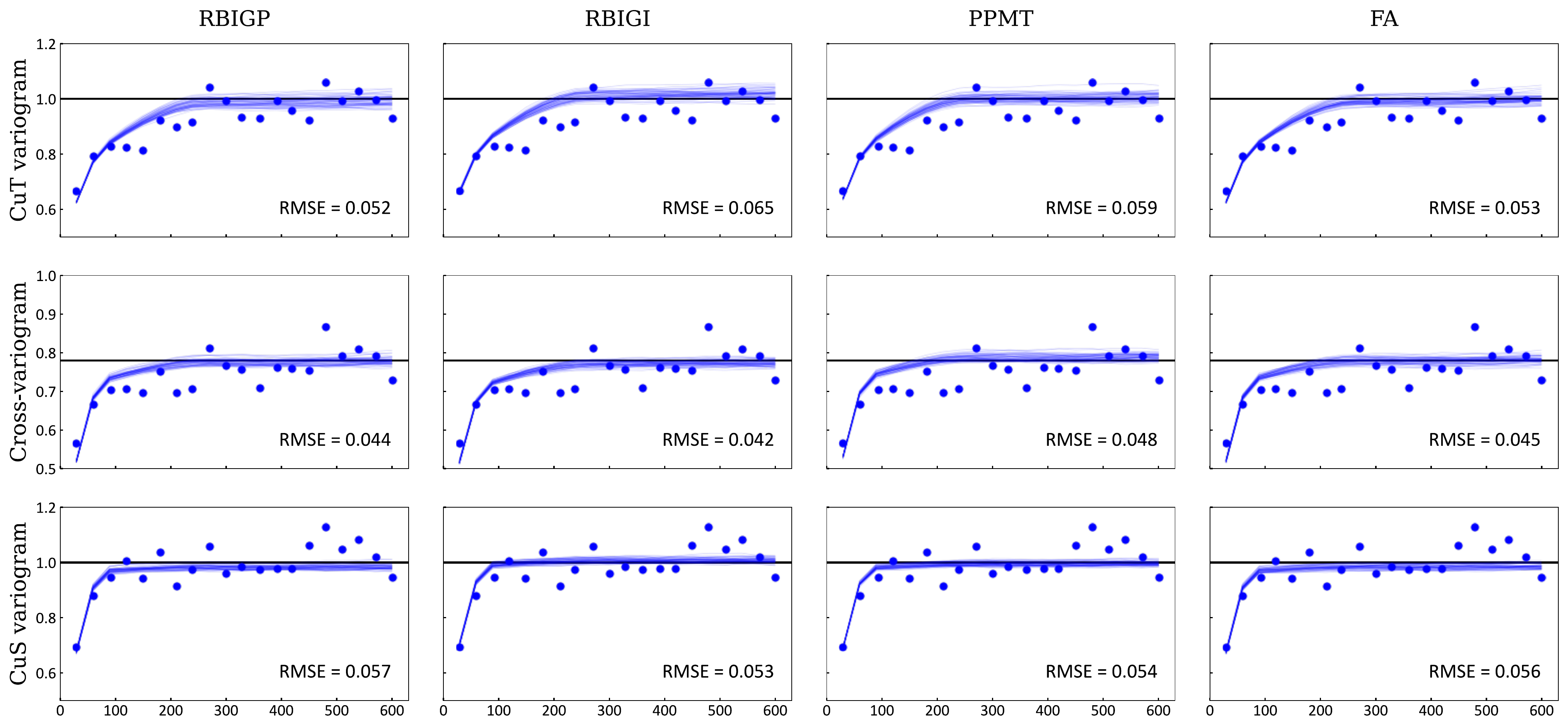}
\caption{Variogram validation for case A with direct variograms in the top and bottom rows and cross-variograms in the middle row. Solid lines represent normal score variograms of simulated realisations and experimental variograms are shown as points.}\label{fig13}
\end{figure}

Figure \ref{fig13} shows the variogram validation for case A. As expected, direct variograms are well reproduced, and cross-variograms of the simulated realisations are also similar to the original points. As was observed in the histogram validation, all three methods show almost identical results from a visual standpoint. Nevertheless, it is difficult to make a comparison based on RMSE results. RBIGP and FA have smaller RMSE for total copper, but RBIGI shows better reproduction for soluble copper and cross-variability. At the same time, RBIGI also has a higher RMSE for total copper.       

\begin{figure}[ht]%
\centering
\includegraphics[width=1\textwidth]{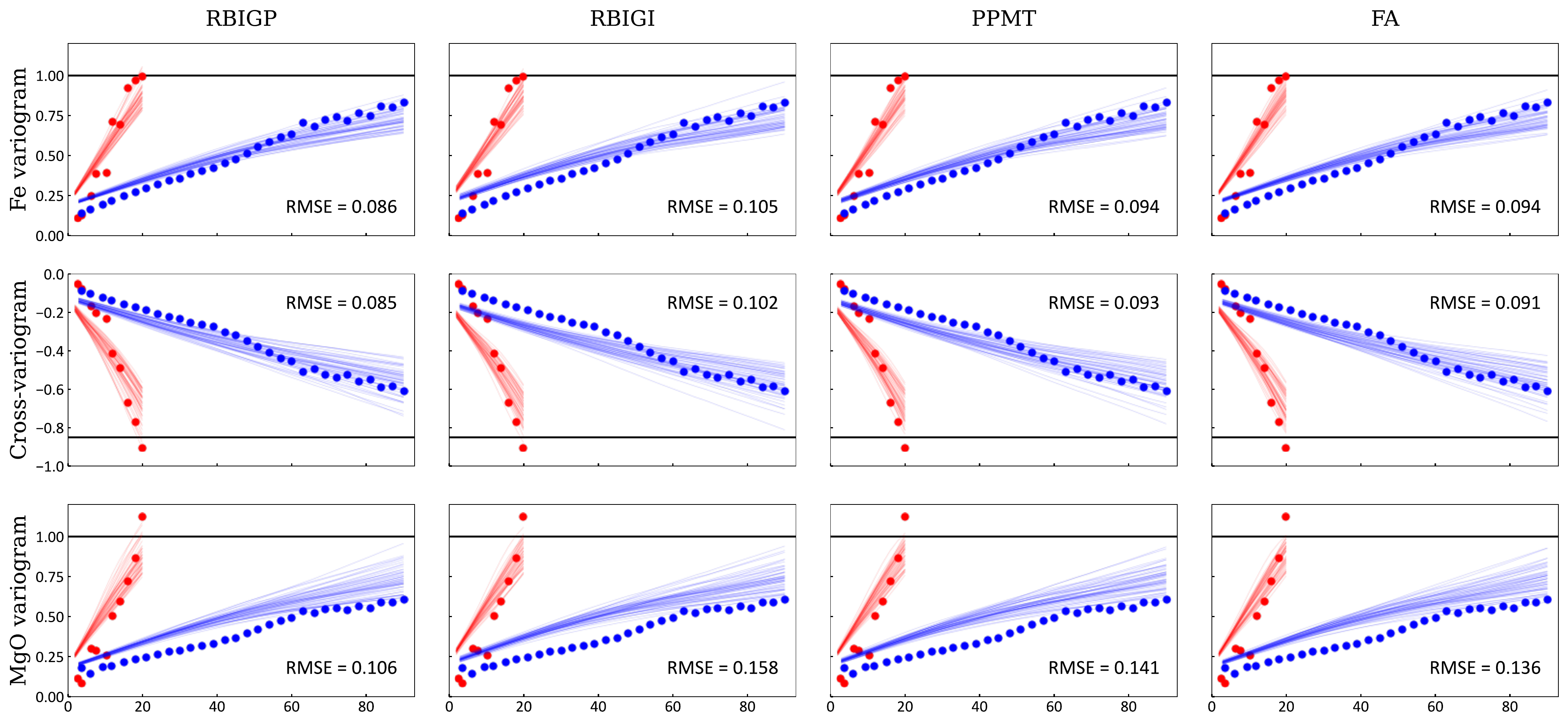}
\caption{Variogram validation for case B with direct variograms in the top and bottom rows and cross-variograms in the middle. Solid lines represent normal score variograms of simulated realisations and experimental variograms are shown as points (blue: horizontal, red: vertical).}\label{fig14}
\end{figure}

In case B, omni-horizontal and vertical variograms were used during the modelling due to anisotropy between those directions. Figure \ref{fig14} shows the variogram validation in both the vertical and horizontal directions. Visually, the results are identical for direct and cross-variograms in both directions. However, based on RMSE, RBIGP performed the best, followed closely by other methods. There is also an underestimation of the magnesium oxide variogram for all three methods.

\begin{figure}[ht]%
\centering
\includegraphics[width=1\textwidth]{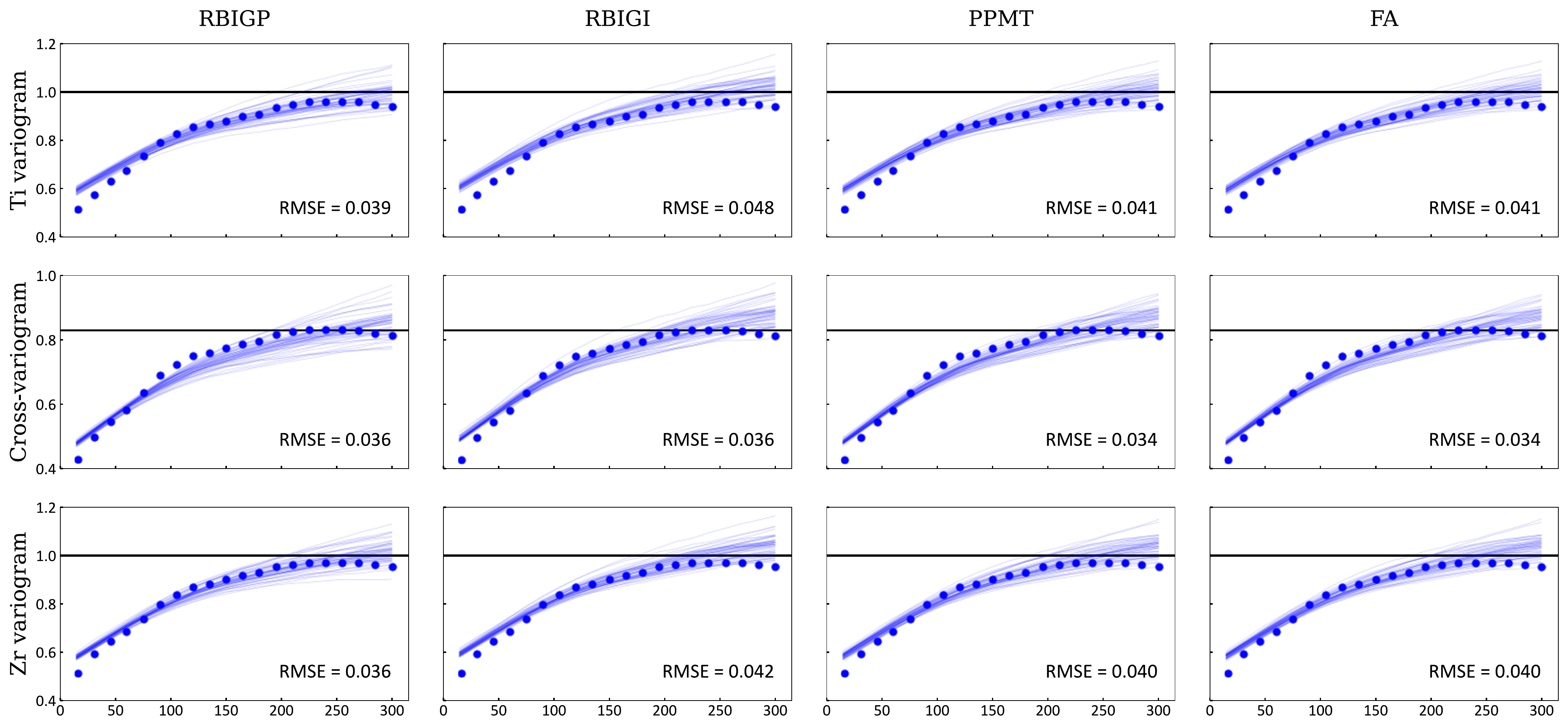}
\caption{Variogram validation for case C with direct variograms in the top and bottom rows and cross-variogram in the middle row. Solid lines represent normal score variograms of simulated realisations and experimental variograms are shown as points.}\label{fig15}
\end{figure}

Finally, variogram reproduction is visually indistinguishable in case C (Figure \ref{fig15}). RBIGP has a smaller RMSE for direct variograms, but the difference with other methods is insignificant. Overall, geostatistical validation metrics suggest that the performance of all four methods is similar, especially in the reproduction of marginal distributions and spatial variability. Moreover, without calculating RMSE, it is impossible to tell the difference between RBIGP, RBIGI, PPMT and FA results.

\section{Discussion and Conclusions}\label{sec4}

This paper compares four MGT methods, PPMT, FA and two types of RBIG (i.e., RBIGP with PCA rotations and RBIGI with ICA rotations), using three case studies with complex multivariate relationships. All three cases were obtained from undisclosed datasets, and each represents a particular bivariate complexity with moderate-to-high correlation coefficients. Case A is a bivariate dataset with an inequality constraint between total and soluble copper grades, with 6,074 data points. It can also be called a fractional constraint since the soluble copper grade is a fraction of the total grade. Case B consists of 9,990 iron and magnesium oxide samples with a non-linear relationship between them. Finally, case C has 33,021 data points, with clear heteroscedasticity between titanium and zirconium grades.

The most apparent difference between MGT approaches in this study is the execution time for forward and back transformation. Although RBIGI, PPMT and FA showed comparable times for forward transformation in cases A and B, FA took almost thrice as much time as PPMT and RBIGI in case C. Furthermore, the difference in execution time is significantly higher during the back transformation. RBIGP, RBIGI and PPMT do not need original data for the inverse transform because they save all the rotation matrices and Gaussian tables at each iteration during the fitting of transforms. Since FA requires the original data, the number of samples of the original hard data significantly affects its execution time. For example, FA required around 14 minutes to back-transform a single realization in cases A and B, while RBIGP, RBIGI and PPMT required only seconds. Moreover, because case C is significantly larger than the other two datasets, the time required by FA to back-transform one realisation was about 35 minutes. Due to a simpler and faster marginal Gaussianisation and more optimised matrix operations in MATLAB, RBIGP was significantly faster than other methods in forward transformation and similar to PPMT in inverse transformation. For example, RBIGP needed less than one second to transform the original data in all three cases.

Nevertheless, FA shows better reproduction of bivariate relationships, which was also observed by other studies reported in the geostatistical literature. PPMT has minor artefacts generated during back-transformation. While RBIGI appears to have good reproduction, it can produce significant outliers from the distribution but still within the bounds of marginals. The reason for these outliers is the convergence issue of ICA rotation, which can be unstable in some iterations. On the other hand, RBIGP does not appear to have this problem, producing results of similar quality to PPMT. Despite the minor differences, all three case studies showed that MGT methods effectively model datasets with multivariate complexities. Furthermore, all three methods show identical results in histogram and variogram validation. Even the RMSE between realisations and the original data shows no significant differences in the RBIGP, RBIGI, PPMT and FA results. Interestingly, the multivariate normality of the multi-Gaussian factors did not significantly impact the results. For example, PPMT failed MVN tests in case B but still produced realisations with a similar quality to those of other methods. However, multivariate normality plays a much bigger role in high-dimensional cases, particularly when correlations between variables are weak. Nevertheless, spatial decorrelation is much more important in geostatistical applications, and PPMT had similar spatial diagonalisation results to those of the other three methods.

Overall, the four MGT approaches analysed in this paper produce similar results in terms of multivariate normality, spatial decorrelation, reproduction of bivariate relationships, histogram and variogram validation. The main difference is the time it takes them to transform the original data and back-transform geostatistical realisations. It can be concluded that FA is not suitable for tasks requiring faster transformation, particularly when working on large datasets. For example, the recently developed rapid resource model updating methods require Gaussian transformation and must be in near real-time \citep{bib48}. Geostatistical decorrelation methods, such as MAF and FA, were applied to rapidly update multiple cross-correlated variables \citep{bib49,bib50}. However, we believe that RBIG and PPMT are more suitable for such tasks, especially when chained with spatial decorrelation. In future work, we will apply RBIG and PPMT in the rapid updating of multiple variables within iron oxide copper-gold deposits.

\backmatter

\bmhead{Author contributions}

Conceptualization: SA, PAD and CX; Methodology: SA; Investigation: SA; Writing - original draft preparation: SA; Writing - review and editing: SA, PAD and CX; Supervision: PAD and CX; Funding acquisition: PAD and CX. All authors read and approved the final manuscript.

\bmhead{Acknowledgments} 

The research reported here was supported by the Australian Research Council Industrial Transformation Training Centre for Integrated Operations for Complex Resources (ARC ITTC IOCR - project number IC190100017) and funded by universities, industry and the Australian Government. The first author also acknowledges the International Association for Mathematical Geosciences for providing a Travel Grant to attend the IAMG 2022 conference and present this study. Finally, we acknowledge Geovariances for providing a demonstration version of Isatis.neo Mining Edition software.

\bmhead{Code availability}

Python package ``\textit{pymgt}” from ARC TC IOCR was used to perform RBIG and PPMT transformations (available at \url{https://github.com/exepulveda/pymgt}). FA was performed using the ``\textit{gmGeostats}” CRAN package (available at \url{https://cran.r-project.org/web/packages/gmGeostats}). For RBIGP, the original MATLAB code was used (available at \url{https://github.com/IPL-UV/rbig_matlab}). MAF, variogram fitting and geostatistical modelling were performed using the Isatis.neo software.

\bmhead{Conflicts of Interest}

The authors declare no conflict of interest.

\bibliography{bibliography}

\begin{thebibliography}{50}
\providecommand{\natexlab}[1]{#1}
\providecommand{\url}[1]{{#1}}
\providecommand{\urlprefix}{URL }
\providecommand{\doi}[1]{\url{https://doi.org/#1}}
\providecommand{\eprint}[2][]{\url{#2}}
 \bibcommenthead

\bibitem[{Abildin et~al(2019)Abildin, Madani, and Topal}]{bib27}
Abildin Y, Madani N, Topal E (2019) {A Hybrid Approach for Joint Simulation of
  Geometallurgical Variables with Inequality Constraint}. Minerals 9(1):24.
  \doi{10.3390/min9010024}

\bibitem[{Abulkhair and Madani(2021)}]{bib30}
Abulkhair S, Madani N (2021) {Assessing heterotopic searching strategy in
  hierarchical cosimulation for modeling the variables with inequality
  constraints}. Comptes Rendus Géoscience 353(1):115--134.
  \doi{10.5802/crgeos.58}

\bibitem[{Almeida and Journel(1994)}]{bib6}
Almeida AS, Journel AG (1994) {Joint simulation of multiple variables with a
  Markov-type coregionalization model}. Mathematical Geology 26(5):565--588.
  \doi{10.1007/BF02089242}

\bibitem[{Barnett et~al(2014)Barnett, Manchuk, and Deutsch}]{bib13}
Barnett RM, Manchuk JG, Deutsch CV (2014) {Projection Pursuit Multivariate
  Transform}. Mathematical Geosciences 46:337--359.
  \doi{10.1007/s11004-013-9497-7}

\bibitem[{Barnett et~al(2016)Barnett, Manchuk, and Deutsch}]{bib16}
Barnett RM, Manchuk JG, Deutsch CV (2016) {The Projection-Pursuit Multivariate
  Transform for Improved Continuous Variable Modeling}. SPE Journal
  21(06):2010--2026. \doi{10.2118/184388-PA}

\bibitem[{Bassani et~al(2018)Bassani, {Coimbra Leite Costa}, and
  Deutsch}]{bib24}
Bassani MAA, {Coimbra Leite Costa} JF, Deutsch CV (2018) {Multivariate
  geostatistical simulation with sum and fraction constraints}. Applied Earth
  Science 127(3):83--93. \doi{10.1080/25726838.2018.1468145}

\bibitem[{Benndorf(2020)}]{bib48}
Benndorf J (2020) Closed Loop Management in Mineral Resource Extraction.
  Springer, Cham

\bibitem[{Davis and Greenes(1983)}]{bib8}
Davis BM, Greenes KA (1983) {Estimation using spatially distributed
  multivariate data: An example with coal quality}. Journal of the
  International Association for Mathematical Geology 15:287--300.
  \doi{10.1007/BF01036071}

\bibitem[{{de Figueiredo} et~al(2021){de Figueiredo}, Schmitz, Lunelli,
  Roisenberg, {de Freitas}, and Grana}]{bib15}
{de Figueiredo} LP, Schmitz T, Lunelli R, et~al (2021) {Direct Multivariate
  Simulation - A stepwise conditional transformation for multivariate
  geostatistical simulation}. Computers \& Geosciences 147:{104,659}.
  \doi{10.1016/j.cageo.2020.104659}

\bibitem[{Desassis and Renard(1973)}]{bib46}
Desassis N, Renard D (1973) {The intrinsic random functions and their
  applications}. Advances in Applied Probability 5(3):439--468.
  \doi{10.2307/1425829}

\bibitem[{Desassis and Renard(2013)}]{bib45}
Desassis N, Renard D (2013) {Automatic Variogram Modeling by Iterative Least
  Squares: Univariate and Multivariate Cases}. Mathematical Geosciences
  45:453--470. \doi{10.1007/s11004-012-9434-1}

\bibitem[{Desbarats and Dimitrakopoulos(2000)}]{bib10}
Desbarats A, Dimitrakopoulos R (2000) {Geostatistical Simulation of
  Regionalized Pore-Size Distributions Using Min/Max Autocorrelation Factors}.
  Mathematical Geology 32:919--942. \doi{10.1023/A:1007570402430}

\bibitem[{Deutsch and Journel(1992)}]{bib31}
Deutsch CV, Journel AG (1992) GSLIB: Geostatistical Software Library and User's
  Guide. Oxford University Press, New York

\bibitem[{Ebner and Henze(2020)}]{bib42}
Ebner B, Henze N (2020) {Tests for multivariate normality—a critical review
  with emphasis on weighted ${L}^2$-statistics}. TEST 29:845--892.
  \doi{10.1007/s11749-020-00740-0}

\bibitem[{Egozcue et~al(2003)Egozcue, Pawlowsky-Glahn, Mateu-Figueras, and
  Barcelo-Vidal}]{bib23}
Egozcue JJ, Pawlowsky-Glahn V, Mateu-Figueras G, et~al (2003) {Reproduction of
  secondary data in projection pursuit transformation}. Mathematical Geology
  35:279--300. \doi{10.1023/A:1023818214614}

\bibitem[{Emery(2008)}]{bib5}
Emery X (2008) {A turning bands program for conditional co-simulation of
  cross-correlated Gaussian random fields}. Computers \& Geosciences
  34(12):1850--1862. \doi{10.1016/j.cageo.2007.10.007}

\bibitem[{Emery(2010)}]{bib44}
Emery X (2010) {Iterative algorithms for fitting a linear model of
  coregionalization}. Computers \& Geosciences 36(9):1150--1160.
  \doi{10.1016/j.cageo.2009.10.007}

\bibitem[{Emery and Lantuéjoul(2006)}]{bib47}
Emery X, Lantuéjoul C (2006) {TBSIM: A computer program for conditional
  simulation of three-dimensional Gaussian random fields via the turning bands
  method}. Computers \& Geosciences 32(10):1615--1628.
  \doi{10.1016/j.cageo.2006.03.001}

\bibitem[{Erten and Deutsch(2021)}]{bib25}
Erten O, Deutsch CV (2021) {Assessment of variogram reproduction in the
  simulation of decorrelated factors}. Stochastic Environmental Research and
  Risk Assessment 35:2583--2604. \doi{10.1007/s00477-021-02005-0}

\bibitem[{Friedman(1987)}]{bib18}
Friedman JH (1987) {Exploratory Projection Pursuit}. Journal of the American
  Statistical Association 82(397):249--266.
  \doi{10.1080/01621459.1987.10478427}

\bibitem[{Henze and Zirkler(1990)}]{bib40}
Henze N, Zirkler B (1990) {A class of invariant consistent tests for
  multivariate normality}. Communications in Statistics - Theory and Methods
  19(10):3595--3617. \doi{10.1080/03610929008830400}

\bibitem[{Hosseini and Asghari(2015)}]{bib28}
Hosseini SA, Asghari O (2015) {Simulation of geometallurgical variables through
  stepwise conditional transformation in Sungun copper deposit, Iran}. Arabian
  Journal of Geosciences 8:3821--3831. \doi{10.1007/s12517-014-1452-5}

\bibitem[{Hosseini and Asghari(2019)}]{bib21}
Hosseini SA, Asghari O (2019) {Multivariate Geostatistical Simulation on
  Block-Support in the Presence of Complex Multivariate Relationships: Iron Ore
  Deposit Case Study}. Natural Resources Research 28:125--144.
  \doi{10.1007/s11053-018-9379-2}

\bibitem[{Hyvarinen(1999)}]{bib32}
Hyvarinen A (1999) {Fast and robust fixed-point algorithms for independent
  component analysis}. IEEE transactions on Neural Networks 10(3):626--634.
  \doi{10.1109/72.761722}

\bibitem[{Joenssen and Vogel(2014)}]{bib43}
Joenssen DW, Vogel J (2014) {A power study of goodness-of-fit tests for
  multivariate normality implemented in R}. Journal of Statistical Computation
  and Simulation 84(5):1055--1078. \doi{10.1080/00949655.2012.739620}

\bibitem[{Journel(1999)}]{bib7}
Journel AG (1999) {Markov models for cross-covariances}. Mathematical Geology
  31(8):955--964. \doi{10.1023/A:1007553013388}

\bibitem[{Journel and Huijbregts(1978)}]{bib1}
Journel AG, Huijbregts CJ (1978) Mining geostatistics. Academic Press, London

\bibitem[{Korkmaz et~al(2014)Korkmaz, Göksülük, and Zararsiz}]{bib37}
Korkmaz S, Göksülük D, Zararsiz G (2014) {MVN: An R Package for Assessing
  Multivariate Normality}. R JOURNAL 6(2):151--162

\bibitem[{Kumar et~al(2020)Kumar, Dimitrakopoulos, and Maulen}]{bib49}
Kumar A, Dimitrakopoulos R, Maulen M (2020) {Adaptive self-learning mechanisms
  for updating short-term production decisions in an industrial mining
  complex}. Journal of Intelligent Manufacturing 31:1795--1811.
  \doi{10.1007/s10845-020-01562-5}

\bibitem[{Laparra et~al(2011)Laparra, Camps-Valls, and Malo}]{bib26}
Laparra V, Camps-Valls G, Malo J (2011) {Iterative Gaussianization: From ICA to
  Random Rotations}. IEEE Transactions on Neural Networks 22(4):537--549.
  \doi{10.1109/TNN.2011.2106511}

\bibitem[{Leuangthong and Deutsch(2003)}]{bib14}
Leuangthong O, Deutsch CV (2003) {Stepwise Conditional Transformation for
  Simulation of Multiple Variables}. Mathematical Geology 35(2):155--173.
  \doi{10.1023/A:1023235505120}

\bibitem[{Madani and Abulkhair(2020)}]{bib29}
Madani N, Abulkhair S (2020) {A hierarchical cosimulation algorithm integrated
  with an acceptance–rejection method for the geostatistical modeling of
  variables with inequality constraints}. Stochastic Environmental Research and
  Risk Assessment 34:1559--1589. \doi{10.1007/s00477-020-01838-5}

\bibitem[{Manchuk et~al(2017)Manchuk, Barnett, and Deutsch}]{bib22}
Manchuk JG, Barnett RM, Deutsch CV (2017) {Reproduction of secondary data in
  projection pursuit transformation}. Stochastic Environmental Research and
  Risk Assessment 31:2585--2605. \doi{10.1007/s00477-016-1363-y}

\bibitem[{Mardia(1970)}]{bib38}
Mardia KV (1970) {Measures of multivariate skewness and kurtosis with
  applications}. Biometrika 57(3):519--530. \doi{10.1093/biomet/57.3.519}

\bibitem[{Mueller and Ferreira(2012)}]{bib11}
Mueller UA, Ferreira J (2012) {The U-WEDGE Transformation Method for
  Multivariate Geostatistical Simulation}. Mathematical Geosciences
  44:427--448. \doi{10.1007/s11004-012-9384-7}

\bibitem[{Mueller et~al(2020)Mueller, Tolosana-Delgado, Grunsky, and
  McKinley}]{bib12}
Mueller UA, Tolosana-Delgado R, Grunsky EC, et~al (2020) {Biplots for
  compositional data derived from generalized joint diagonalization methods}.
  Applied Computing and Geosciences 8:{100,044}.
  \doi{10.1016/j.acags.2020.100044}

\bibitem[{Pawlowsky-Glahn et~al(2015)Pawlowsky-Glahn, Egozcue, and
  Tolosana-Delgado}]{bib20}
Pawlowsky-Glahn V, Egozcue JJ, Tolosana-Delgado R (2015) Modeling and analysis
  of compositional data. John Wiley \& Sons

\bibitem[{Pedregosa et~al(2011)Pedregosa, Varoquaux, Gramfort, Michel, Thirion,
  Grisel, Blondel, Prettenhofer, Weiss, and Dubourg}]{bib33}
Pedregosa F, Varoquaux G, Gramfort A, et~al (2011) {Scikit-learn: Machine
  Learning in Python}. Journal of Machine Learning Research 12:2825--2830

\bibitem[{Prior et~al(2021)Prior, Tolosana-Delgado, {van den Boogaart}, and
  Benndorf}]{bib50}
Prior A, Tolosana-Delgado R, {van den Boogaart} KG, et~al (2021) {Resource
  Model Updating For Compositional Geometallurgical Variables}. Mathematical
  Geosciences 53:945--968. \doi{10.1007/s11004-020-09874-1}

\bibitem[{Rossi and Deutsch(2014)}]{bib3}
Rossi ME, Deutsch CV (2014) Mineral resource estimation. Springer, Dordrecht

\bibitem[{Royston(1983)}]{bib39}
Royston J (1983) {Some Techniques for Assessing Multivarate Normality Based on
  the Shapiro-Wilk W}. Journal of the Royal Statistical Society: Series C
  (Applied Statistics) 32(2):121--133. \doi{10.2307/2347291}

\bibitem[{Székely and Rizzo(2013)}]{bib41}
Székely GJ, Rizzo ML (2013) {Energy statistics: A class of statistics based on
  distances}. Journal of statistical planning and inference 143(8):1249--1272.
  \doi{10.1016/j.jspi.2013.03.018}

\bibitem[{Talebi et~al(2019)Talebi, Mueller, Tolosana-Delgado, and {van den
  Boogaart}}]{bib35}
Talebi H, Mueller U, Tolosana-Delgado R, et~al (2019) {Geostatistical
  Simulation of Geochemical Compositions in the Presence of Multiple Geological
  Units: Application to Mineral Resource Evaluation}. Mathematical Geosciences
  51:129--153. \doi{10.1007/s11004-018-9763-9}

\bibitem[{Tercan and Sohrabian(2013)}]{bib9}
Tercan A, Sohrabian B (2013) {Multivariate geostatistical simulation of coal
  quality data by independent components}. International Journal of Coal
  Geology 112:53--66. \doi{10.1016/j.coal.2012.10.007}

\bibitem[{Tercan(1999)}]{bib51}
Tercan AE (1999) {Importance of Orthogonalization Algorithm in Modeling
  Conditional Distributions by Orthogonal Transformed Indicator Methods}.
  Mathematical Geology 31:155--173. \doi{10.1023/A:1007557701073}

\bibitem[{Tolosana-Delgado and Mueller(2021)}]{bib36}
Tolosana-Delgado R, Mueller U (2021) Geostatistics for Compositional Data with
  R. Springer, Cham

\bibitem[{Tolosana-Delgado et~al(2019)Tolosana-Delgado, Mueller, and {van den
  Boogaart}}]{bib19}
Tolosana-Delgado R, Mueller U, {van den Boogaart} KG (2019) {Geostatistics for
  compositional data: an overview}. Mathematical Geosciences 51(4):485--526.
  \doi{10.1007/s11004-018-9769-3}

\bibitem[{{van den Boogaart} et~al(2017){van den Boogaart}, Mueller, and
  Tolosana-Delgado}]{bib17}
{van den Boogaart} KG, Mueller U, Tolosana-Delgado R (2017) {An Affine
  Equivariant Multivariate Normal Score Transform for Compositional Data}.
  Mathematical Geosciences 49:231--251. \doi{10.1007/s11004-016-9645-y}

\bibitem[{Verly(1993)}]{bib4}
Verly G (1993) {Sequential Gaussian cosimulation: a simulation method
  integrating several types of information}. In: Soares A (ed) Geostatistics
  Tróia ’92. Quantitative Geology and Geostatistics, vol 5. Springer,
  Dordrecht, p 543--554, \doi{10.1007/978-94-011-1739-5_42}

\bibitem[{Wackernagel(2003)}]{bib2}
Wackernagel H (2003) Multivariate Geostatistics: An Introduction with
  Applications. Springer, Berlin

\end{thebibliography}

\end{document}